\documentclass[twocolumn]{svjour3}
\usepackage{graphicx,array,amssymb}
\usepackage{amsfonts,amsmath,mathrsfs,booktabs}
\usepackage{multirow,subfigure,times,epsfig}
\usepackage{threeparttable,chngpage,float,color,bm}
\usepackage{empheq}
\usepackage{epstopdf,hhline}
\usepackage{multicol}
\newlength{\halfpagewidth}
\divide\halfpagewidth by 1
\newcommand{\leftsep}{%
\noindent\raisebox{4mm}[0ex][0ex]{%
\makebox[\halfpagewidth]{\hrulefill}\hbox{\vrule height 3pt}}%
\vspace*{-2mm}%
}
\newcommand{\rightsep}{%
\noindent\hspace*{\halfpagewidth}%
\rlap{\raisebox{-3pt}[0ex][0ex]{\hbox{\vrule height 3pt}}}%
\makebox[\halfpagewidth]{\hrulefill}%
}
\begin{document}
\title{Oscillatory evolution of collective behavior in evolutionary games played with reinforcement learning}
%\title{\color{blue}Periodic evolution in $2\times 2$ game under the reinforcement learning}
\author{Si-Ping Zhang   \and
        Ji-Qiang Zhang  \and
        Li Chen          \and
        Xu-Dong Liu
       }
\institute{
           Si-Ping Zhang \at
           The Key Laboratory of Biomedical Information Engineering of Ministry of Education, The Key Laboratory of Neuro-informatics \& Rehabilitation Engineering of Ministry of Civil Affairs, and Institute of Health and Rehabilitation Science, School of Life Science and Technology, Xi'an Jiaotong University, Xi'an 710049, China
           \and
           Ji-Qiang Zhang \at
           Beijing Advanced Innovation Center for Big Data and Brain Computing, Beihang University, Beijing, 100191, China\\
           zhangjq13@lzu.edu.cn
           \and
           Li Chen \at
           School of Physics and Information Technology, Shaanxi Normal University, Xi'an, 710062, China\\
           \and
           Xu-Dong Liu \at
           Beijing Advanced Innovation Center for Big Data and Brain Computing, Beihang University, Beijing, 100191, China  \\
           }
\maketitle
\begin{abstract}
{Large-scale cooperation underpins the evolution of ecosystems and the human society, and the collective behaviors by self-organization of multi-agent systems are the key for understanding. As artificial intelligence (AI) prevails in almost all branches of science, it would be of great interest to see what new insights of collective behavior could be obtained from a multi-agent AI system. Here, we introduce a typical reinforcement learning (RL) algorithm -- Q learning into evolutionary game dynamics, where agents pursue optimal action on the basis of the introspectiveness rather than the birth-death or imitation processes in the traditional evolutionary game (EG). We investigate the cooperation prevalence numerically for a general $2\times 2$ game setting. We find that the cooperation prevalence in the multi-agent AI is amazingly of equal level as in the  traditional EG in most cases. However, in the snowdrift games with RL we also reveal that explosive cooperation appears in the form of periodic oscillation, and we study the impact of the payoff structure on its emergence. Finally, we show that the periodic oscillation can also be observed in some other EGs with the RL algorithm, such as the rock-paper-scissors game. Our results offer a reference point to understand emergence of cooperation and oscillatory behaviors in nature and society from AI's perspective.
}

\keywords{Self-organization, Artificial intelligence, Evolutionary games, Reinforcement learning, Collective behaviors,
Oscillation, Explosive events}
% \PACS{PACS code1 \and PACS code2 \and more}z
\end{abstract}

\section{Introduction}
\label{introduction}

In the ecosystem and human society, the phenotypic traits of different species and their behavior characters are very complex and diverse
~\cite{N:2006,AJ:2012,KFJ:2016,SA:2007,B:2011,H:1964,RC:2008,DM:2004}, which remain a puzzle till now.
In 1973, the pioneering framework of evolutionary game (EG) theory was proposed to investigate the frequency of the competing populations in the ecosystem by incorporating the classic game theory and the concept of evolution ~\cite{JG:1973}. Inspired by the idea,
many research emerge~\cite{AD:2017,J:2014,JGLJ:2018,JAR:2018,BP:2018,D:2008,A:2007,LBVL:2018,JCMF:2015}, with emphasis on the mechanisms behind the emergence of cooperation among unrelated individuals~\cite{MJDZSA:2017,AD:2016,DW:2009,LCP:2019}, and various mechanisms are revealed, such as direct or indirect reciprocity~\cite{J:2014,AD:2016,CLKM:2017},
topological effect~\cite{JAR:2018,BP:2018,RJE:2016,SM:2017,DMJN:2014,WJGB:2006,PM:2012,CJA:2008}, self-adaption~\cite{BP:2018,WYPP:2014,ZZ:2014}, among others
~\cite{JAR:2018,RJE:2016,SM:2017,RJ:2012,JHEM:2009}.

In parallel, machine learning flourishes in the past decades ~\cite{MCM:2013,YYG:2015,JR:2017,RA:2018,AH:2006,LRBD:2010}
 that facilitates the applications of Artificial Intelligence (AI) to many other fields,
such as pattern recognition \cite{JR:2017,N:2007,FCCNL:2013}, disease prediction~\cite{LXY:2013,BS:2018,PSMDJJDL:2018,TJLB:2014},
decision making in games as well as human-level control~\cite{TJLB:2014,VKDAJ:2015,NT:2018,MW:2019}, and so on~\cite{DM:2015,RD:2016,TJYDJEL:2017,RHS:2013}.
Reinforcement learning (RL) as one of the most powerful machine learning approaches ~\cite{RA:2018,AH:2006,LRBD:2010,MW:2019}, which is rooted in psychology and neuroscience,
 has been widely used to solve the problems in terms of states, actions, rewards, and decision making in various environments through exploratory trials~\cite{RA:2018,DTJIM:2018,AM:2003}.
Commonly used RL algorithms include temporal differences~\cite{RA:2018,S:1988}, dynamic programming~\cite{RA:2018,LRBD:2010,BD:2015,K:1967}, Dyna~\cite{RA:2018},
Sarsa~\cite{RA:2018} and Q-learning~\cite{RA:2018,HAD:2016,H:2010,F:2001}, which have made tremendous progresses and become the most exciting fields in AI. While the expertise of RL and AI in general make them a good candidate for understanding the complex behaviors in ecosystem and society such as cooperation, there is still a lack of such cross-fertilization.

Interesting questions are then raised: \emph{what's the performance in term of cooperation level if AI agents play the game together? And what's the difference between AI-agent systems and the traditional evolutionary games following rules like birth-death or imitation processes mimicking human systems?} Addressing these questions is of paramount importance because clarifying the similarities and difference between AI and human system is the primary step to design human-machine systems, which is the inevitable trend in the future.

Here we investigate the collective behavior of AI agents in
$2\times 2$ reinforcement learning evolutionary games (RLEGs), specifically by means of Q-learning algorithm and compare them to the traditional EG.
In each round, an agent in the population initiates a number of trial games with the rest, all individuals try to maximize their payoff and meanwhile they learn from their experiences. A striking finding is that the cooperation in the RLEGs evolve almost identical level as in the traditional EGs. However, in the snowdrift RLEGs we reveal that explosive cooperation appears in the form of periodic oscillation, which is able to improve the cooperation preference to some extent. Furthermore, different from EGs, the emergence of periodic oscillation is ubiquitous if there is a unique and mixed weak Nash equilibrium in the payoff matrix of the game setting, such as rock-paper-scissors and snowdrift game. Finally, we provide a qualitative explanation for these observations in RLEGs and in particular the impact of the learning parameters on the oscillatory behaviors.

%Q-learning, one paradigm in reinforcement learning, is put forward on the basis of the theory of DP
%to maximizes the expected value of the total reward over all rounds~\cite{BD:2015,HAD:2016,H:2010,F:2001,CP:1992} and its improved version combining with deep learning are widely used for any finite Markov decision process in game fields~\cite{NT:2018,DTJIM:2018,MAF:2002,DM:2008,DACAL:2016,KPB:2006,DJKIA:2017}.
%With the applications, abundant work is emerged on the theory of the topic, the dynamics of learning games~\cite{TJ:2013,YJ:2003} and the influence of parameters on the dynamics~\cite{KPB:2006,MMM:2010,JW:2014}, etc.

\section{Results}
\subsection{Reinforcement learning evolutionary game model}
\label{subsec:model}

We start by introducing our reinforcement learning evolutionary game (RLEG), where
each agent could be in one of $n_{s}$ states within the state set $\mathscr{S}=\{s_{\underline 1}, \cdots s_{\underline {n_{s}}}\}$ and each could take one of $n_a$ actions
from the action set $\mathscr{A}=\{a_{\underline 1}, \cdots a_{\underline{n_{a}}}\}$. In each round $\tau$, a random agent $i$ is chosen in the system as an \emph{initiator}, and
plays a battery of pairwise games with the rest individuals (also called \emph{participants}) with $n_a$-actions ($2\times n_{a}$). Here, the elements in $\mathscr{S}$ and $\mathscr{A}$ are distinguished by their nature (state or action) but in typical cases they are just the same e.g. being cooperation (C) or defection (D). At the end of the round, $i$ gets a reward according to its opponents' action and its own according to a payoff matrix
\begin{eqnarray}\nonumber
\Pi=\left(
\begin{array}{ccc}
\Pi_{a_{\underline 1}a_{\underline 1}}&\cdots & \Pi_{a_{\underline{n_{a}}}a_{\underline 1}}\\
\vdots&\ddots&\vdots\\
\Pi_{a_{\underline 1}a_{\underline{n_{a}}}}&\cdots&\Pi_{a_{\underline {n_{a}}}a_{\underline {n_{a}}}}
\end{array}
\right),\label{pay_matrix_22}
\end{eqnarray}
where $\Pi_{aa^{\prime}}$ denotes $i$'s reward if
agent $i$ with action $a$ is against its opponent with action $a^{\prime}$. Therefore, $i$'s average payoff at round $\tau$ is $\bar{\Pi}_{i}(\tau)=\sum_{j\in\Omega\setminus i}\Pi_{a_{i}a_{j}}(\tau)/(N-1)$, where $\Omega\setminus i$ refer to all agents excluding initiator $i$.

\begin{figure}[htbp]
\centering
\includegraphics[width=\linewidth]{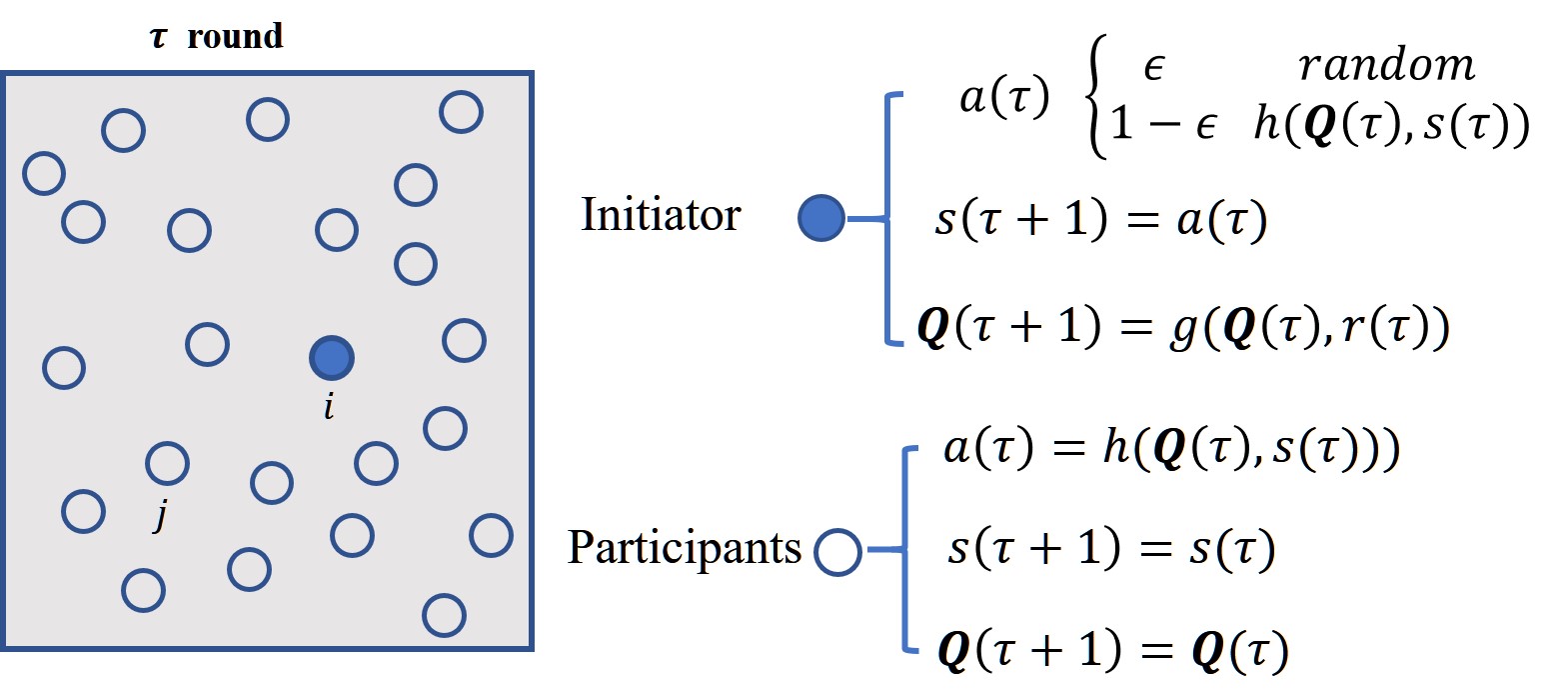}
\caption{ (Color online) {\bf The schematic diagram of the updating process in the well-mixed system.}
At round $\tau$, a game is proposed by an initiator $i$ against all the rest agents known as participants like $j$ in the system. The initiator $i$ takes action according to $h$ function with probability $1-\epsilon$, or acts randomly otherwise. The initiator $i$ both updates its $Q$-table following $g$ function and replaces its state $s$ by current
action at the end of round $\tau$, while a participant $j$ only takes action following its Q-table
in response, but without updating its Q-table and state.}
\label{fig:model}
\end{figure}

In the classical Q-learning algorithm~\cite{CP:1992}, each agent seeks for optimal strategies in the sense that it maximizes the expected values of total reward by
updating the so-called Q-table through learning. A key difference between initiators and participants is that
initiators update both their states and Q-table, while each participant only takes one action as a response, see Fig.\ref{fig:model}.
This setting accounts for the fact that the initiators are actively engaged in the game that they seek for higher rewards, and always try to improve their wisdom via Q-table during the process, while participants are only passively involved in the games proposed without the expectation of lifting wisdom. Note that, the setting is just equivalent to asynchronous updating of Monte Carlo (MC) simulations~\cite{CMA:2005,ZAAHR:2013}.

The Q-table is a matrix of state (rows) -- action (columns) combination $\mathscr{S}\times\mathscr{A}\rightarrow \Re$ as
\begin{eqnarray}\nonumber
\bf{Q}(\tau)=
\left[
\begin{array}{ccc}
Q_{s_{\underline 1}a_{\underline 1}}(\tau)&\cdots&Q_{s_{\underline 1}a_{\underline{n_a}}}(\tau) \\
\vdots&\ddots&\vdots\\
Q_{s_{\underline{n_a}}a_{\underline 1}}(\tau)&\cdots&Q_{s_{\underline{n_s}}a_{\underline{n_a}}}(\tau)
\end{array}
\right].
\end{eqnarray}
With this matrix, if the initiator $i$' state and action are $s$ and
$a$ at $\tau$th round, the element $Q_{sa}$ is updated as follows
\begin{eqnarray}\label{eq:Q_update}
Q_{sa}(\tau+1)&=&g\left({\bf Q}(\tau), r(\tau)\right)\nonumber\\
&=&Q_{sa}(\tau)(1-\alpha)+\alpha\left[r(\tau)+\gamma Q_{s^{\prime}a^{\prime}}^{\max}(\tau)\right],
\label{eq:Q_update}
\end{eqnarray}
where $\alpha\in (0,1]$ is the learning rate and $r(\tau)=\bar{\Pi}_{i}(\tau)$ is the reward received for the action moving from $s(\tau)$ to $s(\tau+1)$. $Q_{s^{\prime}a^{\prime}}^{\max}=\max_{a'}(Q_{s^{\prime}a^{\prime}})$ is the maximum element
in the row of next state $s^{\prime}$, which is the estimate of optimal future value at $s^{\prime}$.
The parameter $\gamma\in[0,1)$ is the discount factor determining the importance of future rewards. Agents with
$\gamma=0$ are short-sighted that they just consider the current rewards, while those with larger $\gamma$
have better foresight. The evolving $Q$-function then can be expressed as ${\bf Q}(\tau+1)=g\left({\bf Q}(\tau),r(\tau)\right)$, and
$Q_{sa}(\tau)$ is replaced by $Q_{sa}(\tau+1)$ at the end of each round.

When the initiator $i$ is in state $s$ at $\tau$th round, it takes action following Q-table
\begin{eqnarray}
a(\tau)=h\left({\bf Q}\left(\tau\right),s(\tau)\right)=
\arg \max\limits_{a^{\prime}}\{Q_{sa^{\prime}}(\tau)\},a^{\prime}\in\mathscr{A}\nonumber
\end{eqnarray}
with probability $1-\epsilon$, or a random action otherwise.
Notice that, $\arg \max\limits_{a^{\prime}}(Q_{sa^{\prime}}(\tau))$ represents the action with the maximal Q-value in the row of state $s$. Each participant in the round always selects the action with the largest value of $Q_{sa^{\prime}}(\tau)$ as Fig.\ref{fig:model} shows. At the end of the round, $i$'s state $s(\tau)$ is instead of $s(\tau+1)=a(\tau)$, i.e. $s^{\prime}=a$ in the Eq.(\ref{eq:Q_update}).

To summarize, the protocol of $Q$-learning
algorithm is as follows:
\begin{itemize}
\item[1)] Initialize matrix $Q$ to zero to mimic the unawareness of agents to the game or environment at beginning,
and initialize state $s$ of each agent randomly.
\item[2)] For each round, a randomly generated
initiator $i$ plays the game with the rest agents and
chooses the action $a$ with largest value of $Q_{sa^{\prime}}(\tau)$ in the
row of current state $s$ with probability $1-\epsilon$, or chooses an
action randomly with probability $\epsilon$.
\item[3)] The $Q_{sa}$ value of the initiator is updated according to Eq.~(\ref{eq:Q_update}),
and the state is also updated as $s(\tau+1) = a(\tau)$.
\item[4)] Each participant only takes the action $a$ with
largest value of $Q_{sa^{\prime}}(\tau)$ in the row of its current state $s$ as response, therefore $s(\tau+1) = s(\tau)$.
\item[5)] Repeat steps 2) -- 4) until the system state becomes statistically stable or evolves to the desired time duration.
\end{itemize}

Notice that agents in RLEGs that adopt reinforcement learning aim to maximize reward gradually rather than birth-death or imitation processes in the traditional evolutionary game and its variants~\cite{N:2006,SA:2007,AJ:2014}. Besides, the coupling between the environment and agents' behaviors gives rise to a dynamic environment, which is different from paradigmatic Q-learning algorithm that individuals confront a static environment. This work is also different from the previously studied minority
game system~\cite{ZZHG:2019} that takes into account a periodic environment.
Our setting then offers a scenario for complex coevolution where the adaption process and environment influence each other and trigger
the emergence of some interesting collective behaviors.
In the following, we will show the main results and the impact of various parameters on the actions and dynamical behaviors in the population.

\subsection{Simulation Results in $2\times 2$ RLEGs}\label{subsec:simulation_results}

In the RLEGs for a $2\times 2$ game setting, the actions set and states set are the same as $\mathscr{A}=\mathscr{S}=\{C, D\}$, and the standard payoff matrix $\Pi=(\Pi_{cc}, \Pi_{cd}; \Pi_{dc}, \Pi_{dd})=(6, b; 6+b, 2)$ with a tunable game parameter $b$. Different ranges of $b$ correspond to different game categories:
1) $b\in[0, 2]$ for Prisoner's Dilemma (PD) with a strict pure Nash equilibrium;
2) $[2, 6]$ for Snowdrift (SD) with a weak mixed Nash equilibrium;
3) $(-\infty,0)$ for Stag Hunt (SH) with two strict pure Nash equilibriums and a weak mixed Nash equilibrium;
and 4) $(6,\infty)$ for mixed stable (MS) with a mixed weak Nash equilibrium ~\cite{N:2006}.
The Q-table here is denoted as ${\bf Q}(\tau)=(Q_{cc}(\tau),Q_{cd}(\tau); Q_{dc}(\tau), Q_{dd}(\tau))$ at round $\tau$.

Before going any further, let's first review the traditional EGs for this $2\times 2$ game, and specifically focus on the stable cooperation preference $f_{c}^{*}$ in the mean-field treatment by replicator dynamics equation (RDE)~\cite{N:2006} of EGs.  Denoting $\Delta\Pi_{:d}=\Pi_{dd}-\Pi_{cd}$ and $\Delta\Pi_{:c}=\Pi_{cc}-\Pi_{dc}$, in case \uppercase\expandafter{\romannumeral1}), when $\Delta\Pi_{:c}\cdot \Delta\Pi_{:d}<0$, $f_{c}=1$ is the stable fixed point if $\Delta\Pi_{:d}>0$ (PD), and $f_{c}=0$ is stable otherwise. In case  \uppercase\expandafter{\romannumeral2}), when $\Delta\Pi_{:c}\cdot \Delta\Pi_{:d}>0$, there is a stable mixed fixed point
$f_{c}=\Delta\Pi_{:d}/(\Delta \Pi_{:d}+\Pi_{:c})$ if $\Delta\Pi_{:c}<0$ (SD or MS), otherwise, the fixed-point is unstable (SH). For the latter, cooperation dominates if the initial preference is greater than the point, otherwise, defection dominates.  In case \uppercase\expandafter{\romannumeral2}), cooperators' and defectors' rewards are identical at the mixed fixed-point.

%{Periodic evolution upon self-organization and symmetry breaking for structure of payoff matrix}
% In the section, we will introduce the mainly simulation results for our reinforce learning game model(RLEG)
% in the well-mixed network and regular
% lattice network. The main results involve three parts: 1) the effect on
% the cooperation preference due to parameters tunning; 2) the emergence of periodic evolution in SD game
% 3) the symmetry breaking caused by the form of payoff matrix.
% So, we arrange the subsections as follows. In Subsec.\ref{}, we provide the cooperation preference in the
% well-mixed network under different parameter combinations
% and compare the results with traditional Nash Equilibrium.

% the influence of the parameters on self-adaption
% and resource allocation in the system with interaction between
% the RL and DD populations only.

We first investigate the cooperation prevalence in our RLEGs, as a function of $b$.
%\textcolor{red}{ within the range $[-7, 9]$, why [-2,8] in Fig. 2?}.
Here we employ the fraction
\begin{eqnarray}
\rho_{c}(t)=\sum_{l=1}^N \delta(a(\tau_{t}^l)-C)/N
\end{eqnarray}
to assess the cooperation preference in the $t$th Monte Carlo (MC) step, which includes $N$ rounds of games denotes as $\tau_{t}^{1},\tau_{t}^{2},\cdots,\tau_{t}^{N}$.
When $\delta=1$ if the initiator's action is $C$ at $\tau_{t}^l$ round, and $\delta=0$ otherwise. With these, the average preference over MC step
$\langle\rho_{c}\rangle$, is used to measures the cooperation preference.

%\textcolor{blue}{The time series is divided into a number of windows
%$t=1, 2,\cdots, \tau/N$ and each one consists of $N$ rounds. Rounds
%in $t$th window are denoted as $\tau_{t}^{1},\tau_{t}^{2},\cdots,\tau_{t}^{N}$.
%At each round, a random agent initiate an identical game with
%the rest agents in well-mixed system.
%We employ the fraction
%$\rho_{c}(t)=\sum_{k=1}^N \delta(a(\tau_{t}^k)-C)/N$
%to assess initiators' cooperation preference in the $t$th window,
%where $\delta=1$ if the initiator's action $a(\tau_{t}^k)$ is $C$, and $\delta=0$ otherwise.
%Then, the average preference defined as
%\begin{eqnarray}
%\langle\rho_{c}\rangle=\sum_{t}\rho_{c}(t)/n_{t}
%\end{eqnarray}
%is able to measure the expected cooperation preference,
%where $n_{t}$ is the number of windows after the transient process. }

\begin{figure}[htbp]
\centering
\includegraphics[width=0.95\linewidth]{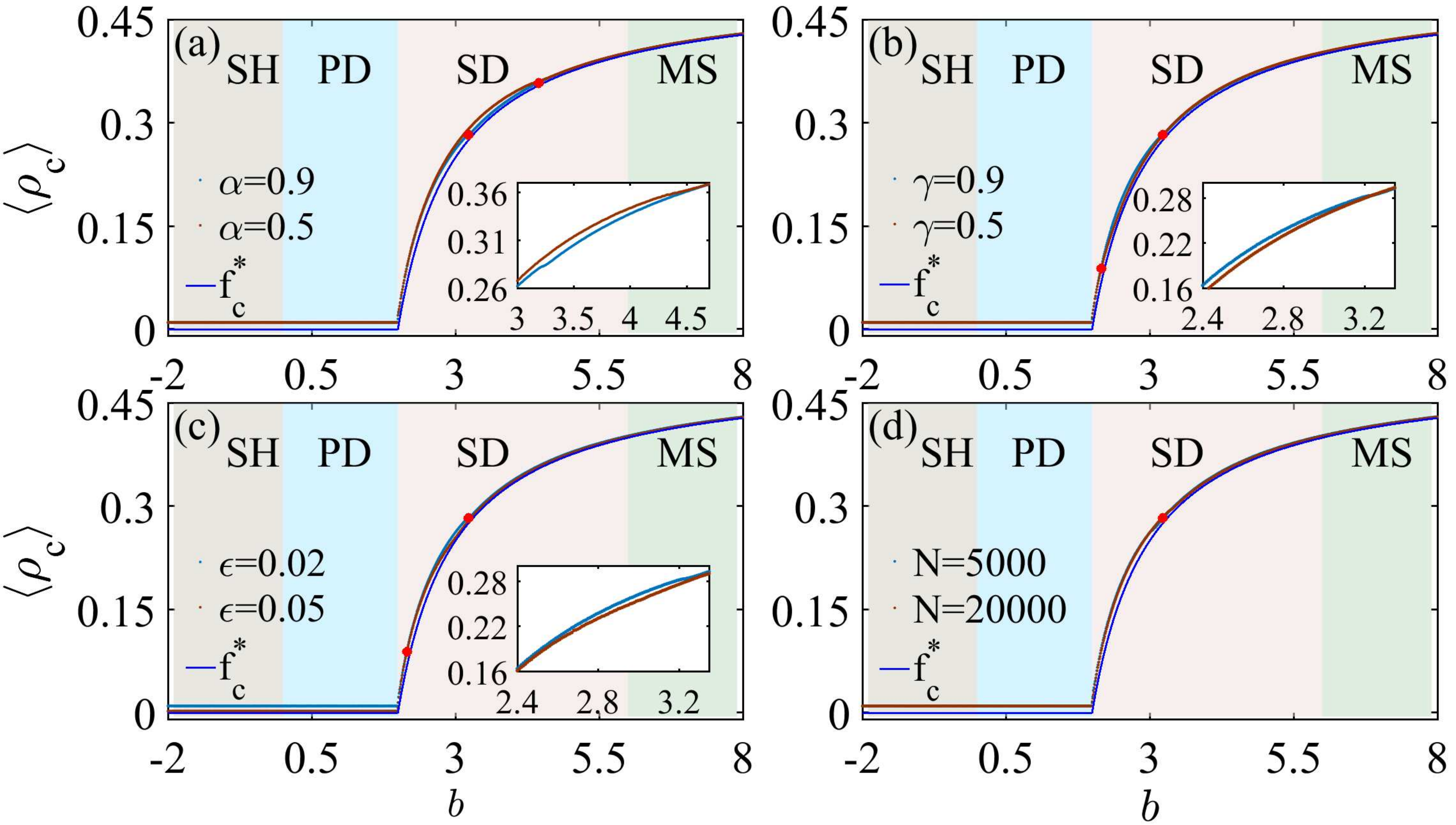}
\caption{ (Color online) {\bf The cooperation preference $\langle\rho_c\rangle$ in RLEGs
 as a function of $b$ under various combination of parameters.}  The cooperation level for traditional EG $f_{c}^*$ is also shown for comparison.
The shared parameters in each panel are as follows: a)  $\gamma=0.9$, $\epsilon=0.02$ and $N=10000$;
b) $\alpha=0.9$, $\epsilon=0.02$ and $N=10000$; c) $\alpha=0.9$, $\gamma=0.9$ and $N=10000$;
d) $\alpha=0.9$, $\gamma=0.9$ and $\epsilon=0.02$. The data is averaged over $n_t=50000$ runs.
}
\label{fig:rhoc_b}
\end{figure}

Fig.~\ref{fig:rhoc_b} shows the cooperation preference both in RLEGs by simulations and in EGs as a function of $b$ for a couple of control parameters.
A striking result is that, in almost all cases $\langle\rho_{c}\rangle$ are close to $f_{c}^*$, which means the cooperation level is equally well when played by AI and by traditional approaches.
And this equivalent performance applies to the whole range of parameters, implying the robustness to specific type of games and parameters.

However, a close lookup shows that $\langle \rho_{c}\rangle$ is slightly greater than $f_{c}^{*}$,
and the difference $\langle \rho_{c}\rangle-f_{c}^{*}$ depends not only on learning parameters, but also on the game category (i.e. different $b$). In SH and PD region, the gap is time-dependent and enlarged with increasing exploration rate $\epsilon$, but is narrowed down when the system approaches into the MS category.
Similar to the dependence of $f_{c}^*$ on initialization in SH EGs,
$\langle \rho_{c}\rangle$ in SH RLEGs also rely on agents' initial Q-table and cooperation preferences (Fig.~S6 in S2.2). Fig.~\ref{fig:rhoc_b}(a-c) show also that the gap is not only sensitive to $\epsilon$ but also to $\alpha$ and $\gamma$ when $b$ is in the SD region.

\begin{figure}[htbp]
\centering
\includegraphics[width=\linewidth]{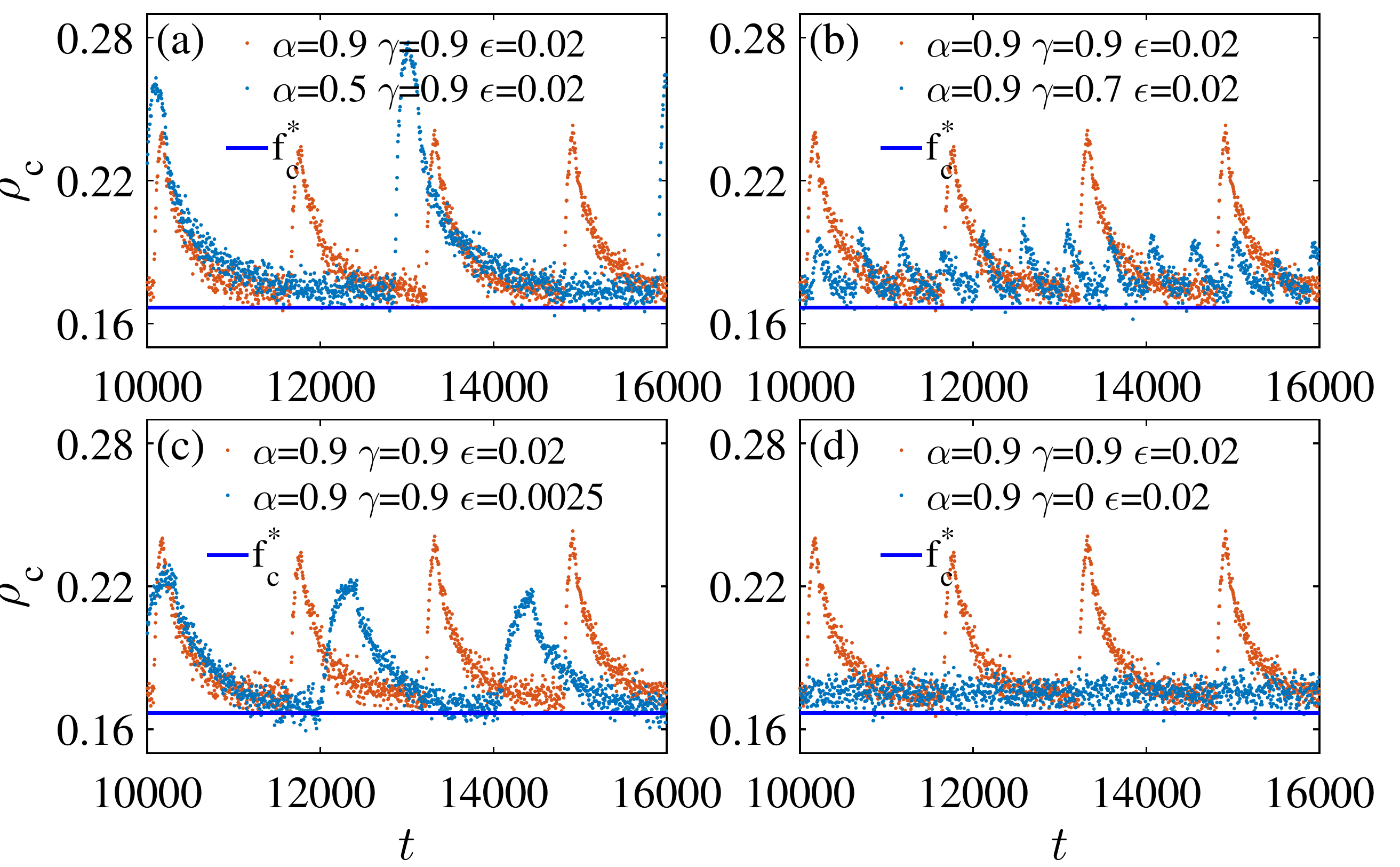}
\caption{ (Color online) {\bf The oscillatory evolution of $\rho_{c}(t)$ under different parameter combinations in RLEGs.}
 The cooperation level for traditional EG $f_{c}^*$ is also shown for reference. The parameters are indicated in each panels}
\label{fig:sd_timeseries}
\end{figure}

To understand these differences, we investigate the time series $\rho_{c}(t)$ for typical cases
and we mainly focus on SD games since their differences are most significant. Shown in Fig.~\ref{fig:sd_timeseries} explains how the difference between $\langle \rho_{c}\rangle$ and $f_{c}^*$ comes from. Unexpectedly, Fig.~\ref{fig:sd_timeseries} reveals an oscillatory structure of $\rho_{c}(t)$ when the games are played by AI agents.
In a typical period, the cooperation preference increases rapidly after a quiescent stage near $f_{c}^*$ and relaxes to $f_{c}^*$ afterwards.
This explains why the cooperation level is enhanced at almost all cases in RLEGs than in the EGs.
Further research shows that the period $T$ and amplitude $A$ of the oscillation increase with $\gamma$ (b),
but decrease with $\alpha$ (a), while larger $\epsilon$ increases the amplitude $A$ but reduces the period $T$ (c).
The oscillation is covered by the noise in the extreme case $\gamma=0$ that all agents are short-sighted as (d) shows.
At last, we discover the oscillatory structure fades away
as $b$ approaches $b^{\prime}$ (Fig.~S5 in S2.1). It implies that the $b^{\prime}$ is the transition point between oscillation and non-oscillation for the cooperation preference.
We would like to note that  in the traditional $2\times2$ EGs, the cooperation preference is always in equilibrium, oscillation is only possible when more than 2 states/actions are available in the system.

\begin{figure}[htbp]
\centering
\includegraphics[width=\linewidth]{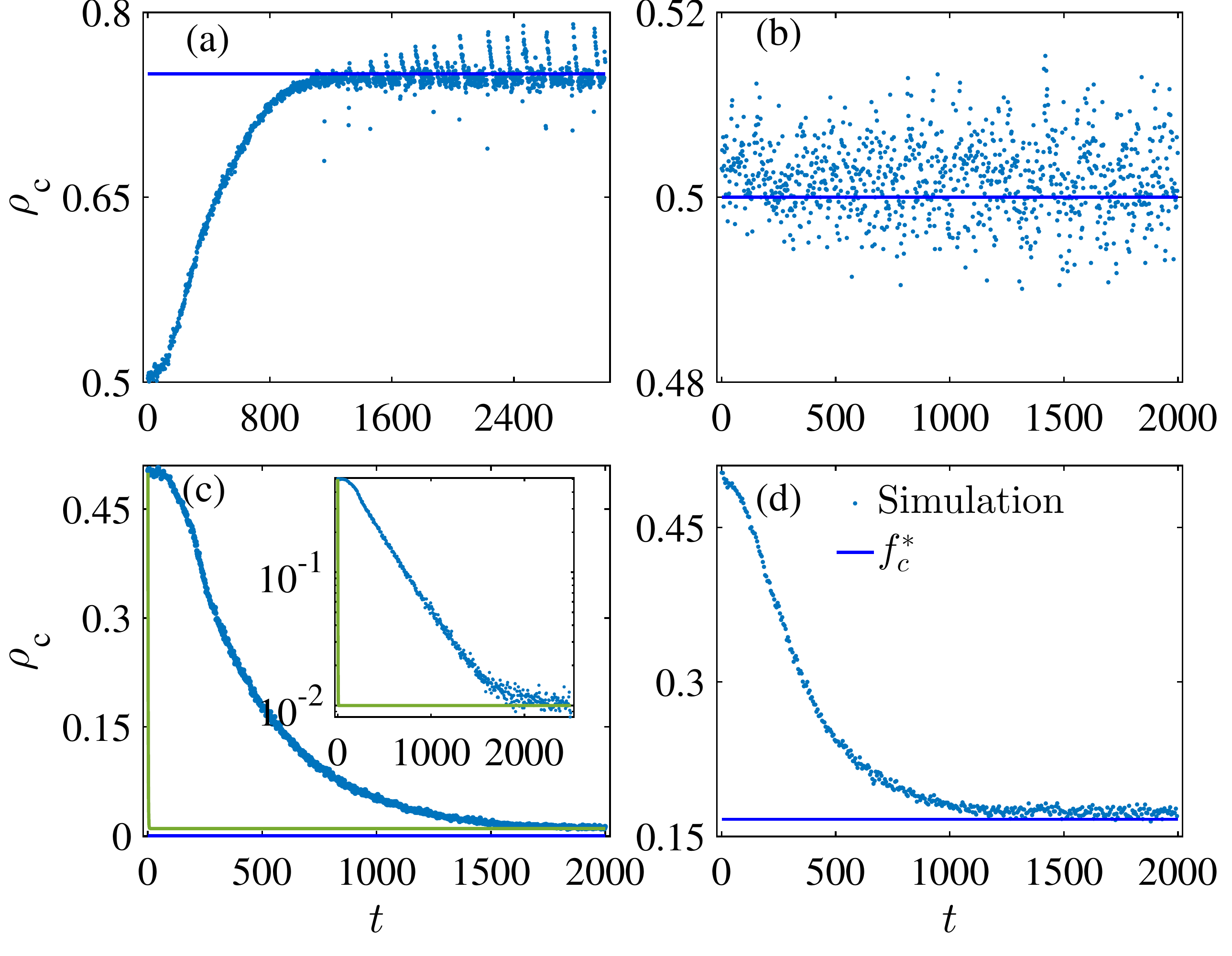}
\caption{ (Color online) {\bf The time series of
$\rho_{c}$ in RLEGs for different payoff structures.} In (a) and (b), the payoff matrices in SD RLEGs
are $\Pi=(6, 5; 7, 2)$ and $(6, 3; 7, 2)$, respectively. (c) shows the time series in a PD RLEG
with $b=1.5$ in a standard payoff matrix to reach the stable value $\rho_{c}^*=\epsilon/2$ . (d) shows the time series in an MS RLEG with the
payoff matrix $(0, 3; 5, 2)$.
The learning parameters $\alpha=\gamma=0.9$, $\epsilon=0.02$
and $N=10000$.}
\label{fig:pd_timeseries}
\end{figure}

Next, we further study the impact of the feature of payoff matrix $\Pi$ on the oscillation dynamics. The above considers the weak Nash equilibrium cases, where is $\Delta\Pi_{:d}/(\Delta \Pi_{:d}+\Pi_{:c})$, is always smaller than $1/2$
in the standard game setting. Here we modify the payoff matrix in SD RLEGs and show that the form of the oscillation is dramatically changed
compared to the scenarios shown in Fig.~\ref{fig:sd_timeseries} when $\Delta\Pi_{:d}/(\Delta \Pi_{:d}+\Pi_{:c})>1/2$ , see Fig.~\ref{fig:pd_timeseries}(a).
In addition, the oscillation fades away as $\Delta\Pi_{:d}/(\Delta \Pi_{:d}+\Pi_{:c})$ tend to $1/2$ as shown in Fig.~\ref{fig:pd_timeseries}(b). These suggest that $\Delta\Pi_{:d}/(\Delta \Pi_{:d}+\Pi_{:c})=1/2$ is the threshold separating oscillation form in the SD RLEGs.
In Fig.~\ref{fig:pd_timeseries}(c-d), we also provide typical time series $\rho_{c}(t)$ for PD and MS RLEGs.
In both case, the cooperation preference $\rho_{c}$ decay with $t$ to the stable level in the end, without any oscillation expected.
These complexities revealed here are absent in the traditional SD EGs, and are unique in multi-agent AI systems.

\section{The analysis}
\subsection{The analysis of the evolution in a static environment}\label{subsec:static_environment}

To understand these observations, we build a system that consists of a number of noninteracting individuals, whose action and state sets $\mathscr{A}\!=\!\mathscr{S}\!=\!\{C, D\}$.
Different from RLEGs, rewards for actions $C$ and $D$ are constant and
denoted as $r_{c}$ and $r_{d}$ respectively, i.e. the environment is assumed to be static for all individuals.
As in RLEG, each individual maximizes reward via reinforcement learning: updates both its Q-table and state.
The updating of Q-table in the Q-learning algorithm involves the self-coupling (memory effect), inter-elements coupling (effect of estimated future reward), and environmental coupling (effect of current reward), through the learning parameters $\alpha$, $\gamma$ and the reward of actions.

\begin{figure}[htbp]
\centering
\includegraphics[width=0.95\linewidth]{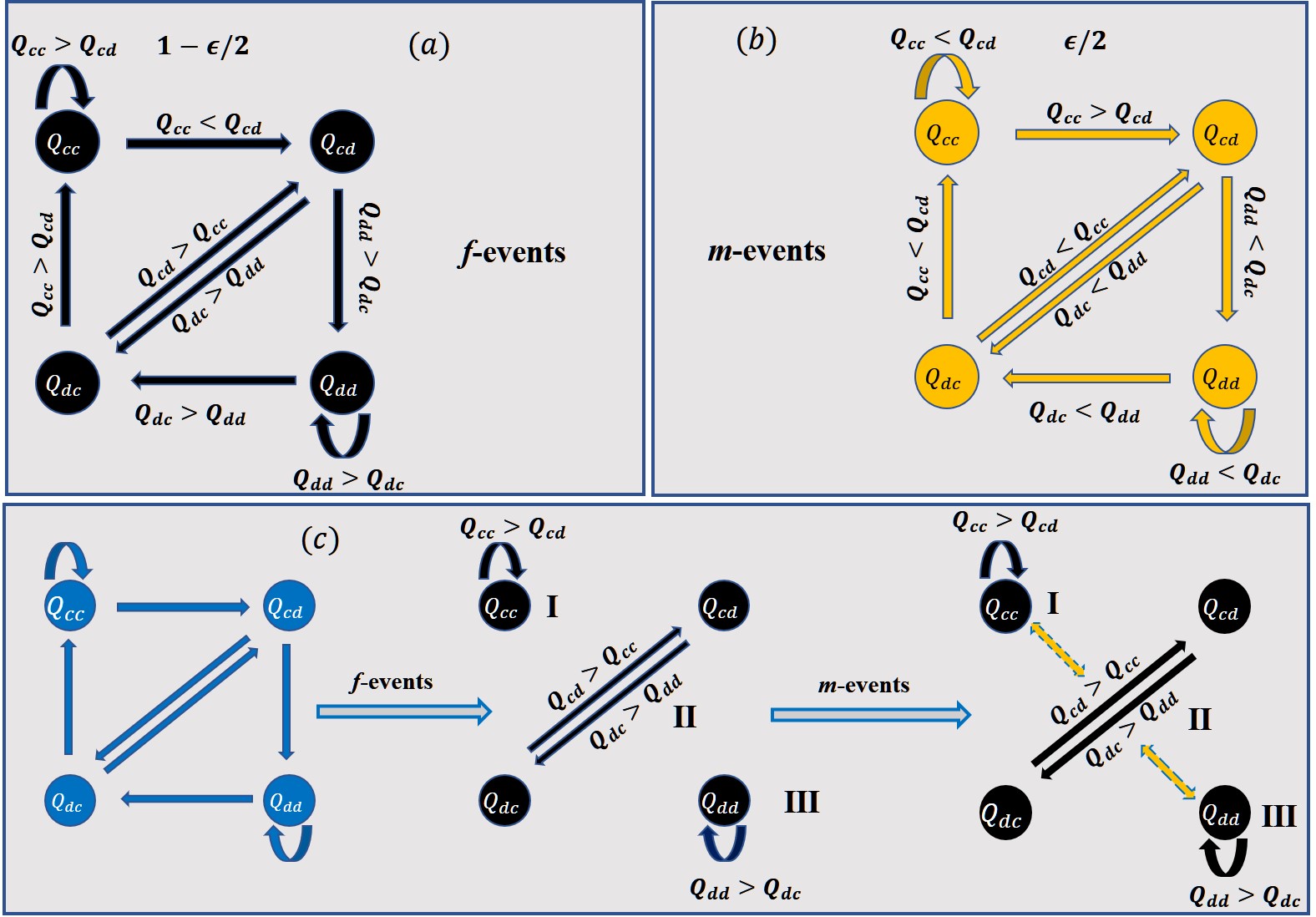}
\caption{ (Color online) {\bf The evolution scheme of Q-table in a static environment.}
(a) and (b) show the hops between elements in Q-table under {\em f}-events and {\em m}-events, respectively.
The conditions of different hops are indicated.
(c) shows the freezing and melting processes of behavioral modes under {\em f}-events and {\em m}-events, respectively. The individual's behavior will be
frozen in one of the three modes (CCM, CDM and DDM) under {\em f}-events. But these modes could be potentially melted and
interchangeable by {\em m}-events.}
\label{fig:Q-function_update}
\end{figure}

To update actions, each individual either follows $h$ function with probability $1-\epsilon$ or acts randomly with $\epsilon$.
Since the random scheme shares half chance with identical actions of following $h$ function, the individual's action updating can then be summarized
as in Fig.~\ref{fig:Q-function_update}(a) and (b).
Fig.~\ref{fig:Q-function_update}(a) shows update hops between elements in the individual's Q-table
if the action is in line with the results following the $h$ function. In contrast,
hops in the cases caused by exploration events is shown in Fig.~\ref{fig:Q-function_update}(b).
The events in Fig.~(a) and (b) could be called as {\it`` freezing events"}({\em f}-events)
and {\it``melting events"}({\em m}-events), respectively.
Because $\epsilon/2\ll 1$ is in the model,
{\em m}-events can be regarded as perturbations in the world of {\em f}-events.

If the system follows only {\em f}-events, each individual's Q-table
will be ``{\em frozen}'' in such a static environment finally.
The freezing rate depends on learning rate $\alpha$ and discounting factor $\gamma$, where a larger $\alpha$ facilitates the freezing process, but $\gamma$ does the opposite (S1.2).
When get frozen, the individuals' behaviors will be in one of three modes as depicted in Fig.~\ref{fig:Q-function_update}(c):
\begin{itemize}
\item[\uppercase\expandafter{\romannumeral1})] frozen cooperation in the form of C-C mode (CCM)
when $\arg \max\limits_{a^{\prime}}\{Q_{sa^{\prime}}\}=C$ for $s=C$;
\item[\uppercase\expandafter{\romannumeral2})] frozen defection in the form of D-D mode (DDM) when $\arg \max\limits_{a^{\prime}}\{Q_{sa^{\prime}}\}=D$ for $s=D$;
\item[\uppercase\expandafter{\romannumeral3})] cyclic frozen mode in the form of cyclic C-D mode (CDM)
when $\arg \max\limits_{a^{\prime}}\{Q_{sa^{\prime}}\}=D$ for $s=C$,
and $\arg \max\limits_{a^{\prime}}\{Q_{sa^{\prime}}\}=C$ for $s=D$.
\end{itemize}
These frozen modes are reminiscent of various attractors in nonlinear dynamics, just the convergence rate towards these attractors is determined by the learning parameters $\alpha$, $\gamma$ and rewards of actions (S1.2), rather than the parameters in equations.

However, when {\em m}-events are present, they do not only perturb individual's actions but may also ``{\it melt}'' modes as Fig.~\ref{fig:Q-function_update}(c) shows. Here, the mode is {\em stable} if it is robust to {\em m}-events that cannot be melted. According to the analysis, we find that only DDM is stable in a static environment as $r_{c}<r_{d}$ (S1.3),
and there is a mode flow CCM $\rightarrow$ CDM $\rightarrow$ DDM if $r_{c}<r_{d}$ and $\epsilon\rightarrow 0$, i.e.
the optimal action is reachable through the reinforcement learning in a static environment.
The dependence of melting rate of the two unstable modes (i.e., CCM and CDM) on $\epsilon$,
$\gamma$, the reward gap $\Delta r=|r_{c}-r_{d}|$, and $\alpha$ shows that the former three factors accelerate the melting process, while $\gamma$
decreases the process (see S1.3). At last, elements of individuals' Q-table in various modes tend to be identical as $r_{c}=r_{d}$.

Intuitively, Q-table are beliefs aiming for the optimal action for individuals in different states and these beliefs are updated gradually in the evolution. Low learning rate means a strong memory effect for individuals because the old Q-table accounts for history takes a large fraction in the updating. As a result, a large $\alpha$ slows down both the freezing and melting processes. However, short-sighted individuals pay much attention to the reward in current round rather than in future, which means these individuals are more sensitive to the reward gap in the exploratory trials. Meanwhile, the probability of trials is positive to the exploration rate $\epsilon$, therefore, $\gamma$, $\epsilon$ and $\Delta r$ accelerate the melting process. Furthermore, a great difference between elements in the row of state $s$ signifies that it takes longer time to change its belief, i.e. the robustness of the beliefs are enslaved to the gaps between elements in the rows.

\subsection{The stability of the $2\times 2$ RLEGs}

In a well-mixed RLEG, the reward for a cooperative or defective initiator at round $\tau$ and $r_{c}(\tau)= p_{c}(\tau)\Delta\Pi_{c:}+\Pi_{cd}$ and $r_{d}(\tau)= -p_{c}(\tau)\Delta\Pi_{d:}+\Pi_{dd}$ as $N\rightarrow \infty$, respectively. Here, $\Delta \Pi_{c:}=\Pi_{cc}-\Pi_{cd}$ and $\Delta \Pi_{d:}=\Pi_{dd}-\Pi_{dc}$, and ${\bf p}(\tau)=\{p_{c}, p_{d}\}$ denotes the shares of cooperators and defectors in the agents following $h$ function, acting as the environment for current initiator. The process to explore the maximal reward is to narrow reward difference between agents in the evolution. Here, we denote the equilibrium as ${\bf p}^*=\{p_{c}^{*}, p_{d}^*\}$, at which the agent is unable to explore a better mode than the current one, therefore is where the evolution of ${\bf p}$ ideally settles down. Analogy to the Nash equilibrium point in $2\times 2$ game, our equilibrium ${\bf p}^*$ can also be classified into two categories: 1) pure and strict, and 2) mixed and weak. For the former, reward for initiators under $f$-events is greater than under $m$-events and $p_{c}^*\in\{0, 1\}$, while for the latter, reward for initiators in $f$-events equal to in $m$-events, i.e. $r_{c}^*=r_{d}^*$ at $p_{c}^*=\Delta\Pi_{:d}/(\Delta \Pi_{:d}+\Pi_{:c})\in(0, 1)$.

However, an equilibrium point ${\bf p}^*$ is stable in a RLEG only if the system is able to resist various fluctuations around ${\bf p}^*$. For the strict case, initiators are unable to be better off by updating their mode at ${\bf p}^*$, such as DDM in PD RLEGs, therefore ${\bf p}^*$ is stable. Notice that each pure equilibrium point is stable in $2\times 2$ RLEGs, e.g. $p_{c}^*=1$ and $p_{c}^*=0$ in the RLEGs for SH setting (Fig.~S7 and S2.2).

For the mixed case, based on the payoff matrix $\Pi$, the standard game settings are further divided into three classes:
\begin{itemize}
\item[\romannumeral1)] $\Delta \Pi_{c:}<0$ and $\Delta \Pi_{d:}<0$ (MS);
\item[\romannumeral2)] $\Delta \Pi_{c:}>0$ and $\Delta \Pi_{d:}>0$ (SH);
\item[\romannumeral3)] $\Delta \Pi_{c:}>0$ and $\Delta \Pi_{d:}<0$ (SD).
\end{itemize}
 The results in the static environment show that the gap between elements in each row decreases with time if reward for initiators as cooperator and defector is identical. Thus, agents' beliefs become fragile gradually to fluctuations regarding ${\bf p}^*$, in which $r_{c}^*=r_{d}^*$. The stability of ${\bf p}^*$ rest with feedback from the change of initiators' belief to various fluctuations, $\delta p_{c}(\tau)=p_{c}(\tau)-p_{c}^*\rightarrow 0$.

For class \romannumeral1), since $r_{c}$ ($r_{d}$) decreases (increases) with $p_{c}$, a fluctuation with $\delta p_{c}(\tau)<0$ cause $r_{c}(\tau)>r_{c}^*$ ($r_{d}(\tau)<r_{d}^*$). The reward change strengthen the beliefs to cooperation but weaken to defection in any events. With the change of the latter belief, $p_{c}$ is increased due to the flux from defectors to cooperators. For a fluctuations with $\delta p_{c}(\tau)>0$, the environment favors defectors rather than cooperators because $r_{c}(\tau)<r_{c}^*$ ($r_{d}(\tau)>r_{d}^*$) in the fluctuation. Hence, $p_{c}$ would decrease in this case. In short, the equilibrium point ${\bf p}^*$ is stable in the MS RLEGs because $\delta p_{c}\cdot dp_{c}/dt<0$ (Fig.\ref{fig:pd_timeseries}(c)). By contrast, in case \romannumeral2) ${\bf p}^*$ is unstable (see Fig.~S6 and S2.2) because the feedback of agent to fluctuations regarding ${\bf p}^*$ in the RLEG is just opposite to class \romannumeral1).

In case \romannumeral3), a standard SD RLEG, both $r_{c}$ and $r_{d}$ increase with $p_{c}$ but the rate of change for $r_{d}$ is higher than for $r_{c}$. Thus, as $\delta p_{c}(\tau)<0$, the beliefs to defection are more fragile than to cooperation in {\em f}-events since $r_{d}(\tau)<r_{c}(\tau)<r_{c}^*=r_{d}^*$ under the fluctuation. Besides, the shares at equilibrium point meet $p_{c}^*<p_{d}^*$. For these, the flux from defectors to cooperators is higher than the opposite, which results in the increase of $p_{c}$. Therefore, the system is able to resist the fluctuations and keep the system around $p_{c}^{*}$ (Fig.~\ref{fig:sd_timeseries}). For $\delta p_{c}(\tau)>0$, the rewards for cooperation and defection are $r_{d}(\tau)>r_{c}^*$ and $r_{c}(\tau)>r_{d}^*$ but $r_{c}(\tau)<r_{d}(\tau)$. So, the beliefs to defection are more fragile than to defection in {\em m}-events. Yet, the flux from defectors to cooperators is still higher than the opposite potentially because $p_{c}^*<p_{d}^*$. The flux will give rise to further increase of $p_{c}$. Hence, the system is unstable for this type of fluctuation. Furthermore, the increase of $\delta p_{c}$ will cause further destruction of agents' beliefs in {\em m}-events. Therefore, an explosive increase of $p_{c}$ is triggered by the cascading effects as Fig.\ref{fig:sd_timeseries} (a-c) shows. The qualitative investigation also indicate the oscillation fades away with the gap $p_{d}^*-p_{c}^*$ (Fig.\ref{fig:rhoc_b}, \ref{fig:pd_timeseries} (b) and Fig.~S6). Moreover, irrhythmic oscillations in non-standard SD RLEGs replace periodic oscillation in standard SD RLEGs if $p_{c}^*>p_{d}^*$ (Fig.~\ref{fig:pd_timeseries} (a) and Fig.~S6).

In fact, during the quiescent stage in standard SD RLEGs, $p_{c}=p_{c}^*$, agents' beliefs are weakened gradually. However, the fluctuations $\delta p_{c}>0$ cause the reward for both actions in the current round is higher than in the past. Therefore, some agents beliefs will be replaced by the new one in {\em m}-events, which triggers the increase of cooperators. Thereby, $\epsilon$ improve rate of increase for $\rho_{c}$ during the explosive stage (Fig.~\ref{fig:sd_timeseries}(c)). In addition, the new belief is strengthened gradually by {\em f}-events due to memory effect at quiescent stage although $r_{c}<r_{d}$ at current stage. Yet, the belief for some cooperators will be melted finally because $r_{c}$ is always lower than $r_{d}$ before $p_{c}$ returns to $p_{c}^*$. Short memory effect and short-sighted means the reward gap takes a greater effect on the freezing as well as melting process as mentioned before. Therefore, high $\alpha$ and low $\gamma$ decrease the period $T$ as well as the amplitude $A$ (Fig.~\ref{fig:sd_timeseries}(a-b)). A higher exploration rate is able to cause a sharply explosive increase of $p_{c}$ but also shorten the melting process. Thus, $\epsilon$ takes a more complex effects to $T$ and $A$ in the oscillation as Fig.~\ref{fig:sd_timeseries}(c) shows.

\subsection{The mean-field analysis of PD RLEGs}\label{subsec:mean_field}
Here we attempt to construct a mathematical framework on the PD RLEG for the mean-field treatment, since there is a single pure strict equilibrium point for the PD game setting. In the standard PD game setting, an arbitrary initiator's reward is $r_{d}(\tau)=\bar{\Pi}_{d}(\tau)=(4+b)p_{c}(\tau)+2$ and $r_{c}(\tau)=\bar{\Pi}_{c}(\tau)=(6-b)p_{c}(\tau)+b$ as a defector and cooperator,
respectively. Here, $r_{d}$ is alway higher than $r_{c}$ due to $p_{c}(\tau)\in[0,1]$. Therefore, the cooperation preference of initiators decreases in exploration trials. And, the preference is the same for initiators and participants just switch their roles in different rounds. This leads to self-consistent results in the end that $p_{c}(\infty)=p_{c}^*=0$ and $\langle\rho_{c}\rangle=\epsilon/2$ because $\langle\rho_{c}\rangle=p_{c}^*(1-\epsilon)+\epsilon/2$. It means that $\langle\rho_{c}\rangle$ only depends on the exploration rate $\epsilon$ in the PD RLEGs (Fig.~\ref{fig:rhoc_b}).
\begin{figure}[htbp]
\centering
\includegraphics[width=0.85\linewidth]{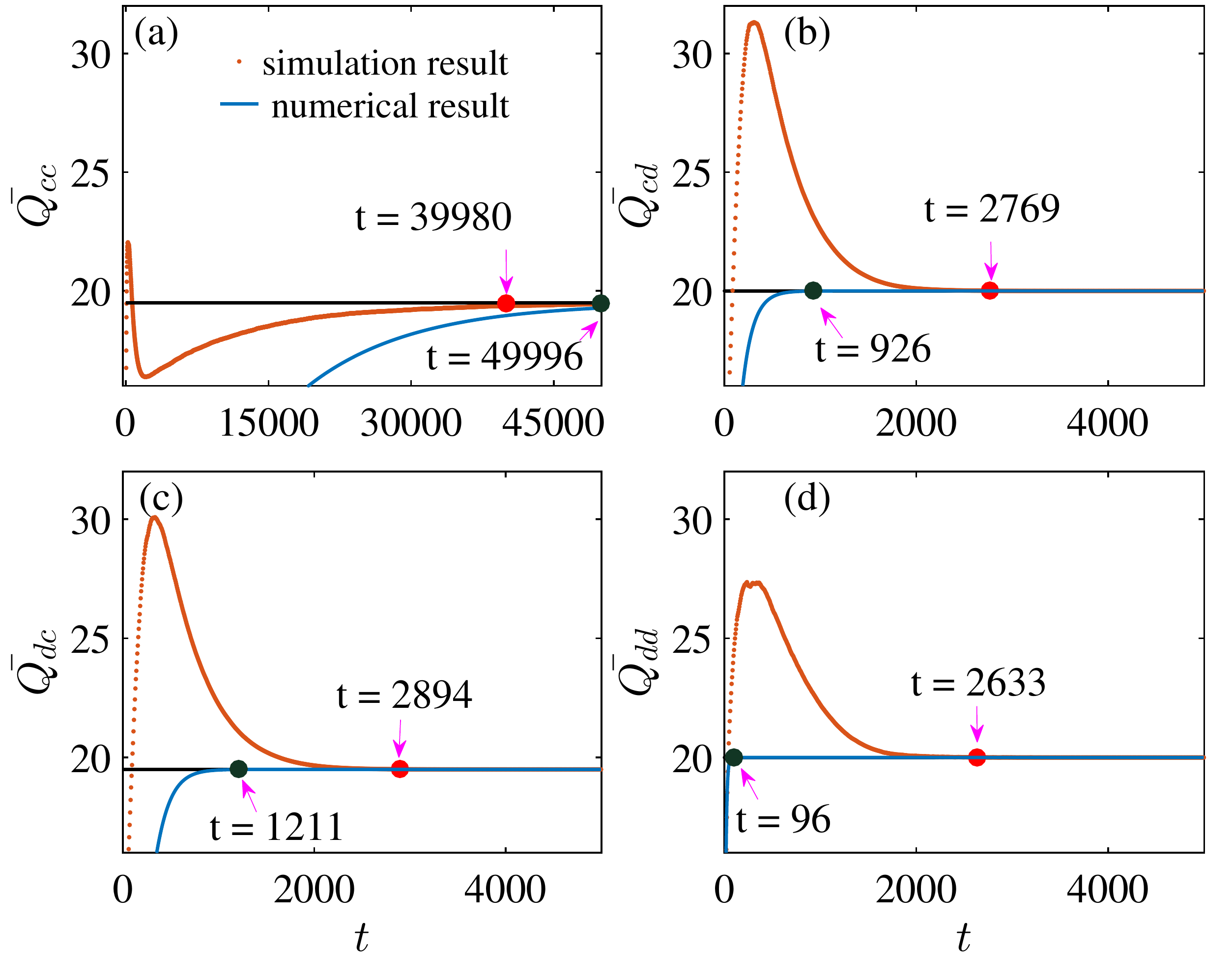}
\caption{ (Color online) {\bf The time evolution of $\bar{{\bf Q}}(t)$
and Q-table $\tilde{\bf Q}^*$ at the stable point $p_{c}^*$.} (a-d) show
the $\bar{Q}_{cc}(t)$, $\bar{Q}_{cd}(t)$, $\bar{Q}_{dc}(t)$ and
$\bar{Q}_{dd}(t)$, respectively. The time points at which the elements become stable
are indicated. The elements in $\tilde{\bf{Q}}^{*}$ are displayed (black
lines) for reference. The parameter setting is the same as Fig.~\ref{fig:pd_timeseries}(a).}
\label{fig:pd_Q-function}
\end{figure}

Thus, the stable Q-table for all agents are uniform and is
\begin{eqnarray}
\tilde{\bf{Q}}^{*}=\left(
\begin{array}{cc}
\frac{\gamma(2-b)+b}{1-\gamma}& \frac{b}{1-\gamma}\\
\frac{\gamma(2-b)+b}{1-\gamma}& \frac{b}{1-\gamma}
\end{array}
\right),
\end{eqnarray}
since the $r_{c}=b$ and $r_{d}=2$ at $p_{c}^*=0$.
To check the analysis, we show the time series of the average Q-table $\bar{\bf{Q}}(t)$
\begin{eqnarray}
\bar{\bf{Q}}(t)=
\left(
\begin{array}{cc}
\frac{\sum_{k=1}^{N}Q_{cc}^{k}(\tau_t^N)}{N} & \frac{\sum_{k=1}^{N}Q_{cd}^{k}(\tau_t^N)}{N}\\
\frac{\sum_{k=1}^{N}Q_{dc}^{k}(\tau_t^N)}{N} & \frac{\sum_{k=1}^{N}Q_{dd}^{k}(\tau_t^N)}{N}
\end{array}
\right)\nonumber
\end{eqnarray}
in the Fig.~\ref{fig:pd_Q-function}. Here, $\tau_t^N$ is
the last round in $t$th MC step and $Q_{sa}^{k}(\tau_{t}^{N})$
denotes the corresponding element of agent $k$'s Q-table at $\tau_{t}^{N}$th round.
$\bar{\bf{Q}}(t)$ and $\tilde{\bf{Q}}^{*}$ in Fig.~\ref{fig:pd_Q-function} show that the result of
stable Q-table is reliable in our analysis.
Moreover, another result in the static environment is also reflected in PD RLEGs, that is the convergence rate for the elements in Q-table decreases as
$\bar{Q}_{dd}$, $\bar{Q}_{cd}$, $\bar{Q}_{dc}$, and $\bar{Q}_{cc}$.
$\tilde {\bf Q}$ manifests one element dominant another one for the rival elements because
$\tilde{Q}_{dd}^{*}>\tilde{Q}_{dc}^{*}$ and $\tilde{Q}_{cd}^{*}>\tilde{Q}_{cc}^{*}$.

A mean-field method is employed to calculate ${\bf Q}(t)$ and $\rho_{c}(t)$ numerically as Fig.~\ref{fig:pd_timeseries}(a) and Fig.~\ref{fig:pd_Q-function} shows.
In Eq.~(\ref{eq:numerical_equation}) of mean-field method, we replace each agents' Q-table with the mean Q-table. Besides, $\rho_{c}(\tau)$ means the fraction of agents at $C$ state in the system, which is slightly different from cases in the simulation.
\begin{figure*}[!htbp]
\begin{multicols}{1}
\end{multicols}
%Bla
\leftsep
\begin{eqnarray}\label{eq:numerical_equation}
\left\{
\begin{array}{rl}
r_{c}(\tau)=&\left[\rho_c(\tau)\cdot\Theta(\Delta\bar{Q}_{c:}(\tau))+\rho_{d}(\tau)\cdot\Theta(-\Delta\bar{Q}_{d:}(\tau))\right]\cdot\Pi_{cc}+\\
&\left[\rho_c(\tau)\cdot\Theta(-\Delta\bar{Q}_{c:}(\tau))+\rho_{d}(\tau)\cdot\Theta(\Delta\bar{Q}_{d:}(\tau))\right]\cdot\Pi_{cd}\\
r_{d}(\tau)=&\left[\rho_c(\tau)\cdot\Theta(\Delta\bar{Q}_{c:}(\tau))+\rho_{d}(\tau)\cdot\Theta(-\Delta\bar{Q}_{d:}(\tau))\right]\cdot\Pi_{dc}+\\
&\left[\rho_c(\tau)\cdot\Theta(-\Delta\bar{Q}_{c:}(\tau))+\rho_{d}(\tau)\cdot\Theta(\Delta\bar{Q}_{d:}(\tau))\right]\cdot\Pi_{dd}\\
\rho_{c}(\tau+1)=&\left\{\rho_{c}(\tau)\cdot\left[\frac{\epsilon}{2}+\left(1-\epsilon)\cdot\Theta(\Delta\bar{Q}_{c:}(\tau)\right)\right]+\rho_{d}(\tau)\cdot\left[\frac{\epsilon}{2}+\left(1-\epsilon)\cdot\Theta(\Delta\bar{Q}_{d:}(\tau)\right)\right]\right\}\cdot\rho_c(\tau) \\
&+\rho_{c}(\tau)\cdot\left[\frac{\epsilon}{2}+(1-\epsilon)\cdot\Theta(-\Delta\bar{Q}_{c:}(\tau))\right]\frac{N\cdot\rho_{c}(\tau)-1}{N}\\
&+\rho_{d}(\tau)\cdot\left[\frac{\epsilon}{2}+(1-\epsilon)\cdot\Theta(-\Delta\bar{Q}_{d:}(\tau)) \right]\frac{N\cdot\rho_{c}(\tau)+1}{N}\\
\bar{Q}_{cc}(\tau+1)=&\rho_c(\tau)\cdot\left[\frac{\epsilon}{2}+(1-\epsilon)\cdot\Theta(\Delta\bar{Q}_{c:}(\tau))\right]\cdot\left[\frac{
(N-1)\cdot\bar{Q}_{cc}(\tau)+(1-\alpha)\cdot\bar{Q}_{cc}(\tau)+\alpha\cdot(\gamma\cdot\bar{Q}^{\max}_{ca^{\prime}}(\tau)+r_{c}(\tau))}{N}\right]\\
&+\left\{1-\rho_{c}(\tau)\cdot\left[\frac{\epsilon}{2}+(1-\epsilon)\cdot\Theta(\Delta\bar{Q}_{c:}(\tau))\right]\right\}\cdot\bar{Q}_{cc}(\tau)\\
\bar{Q}_{dd}(\tau+1)=&\rho_d(\tau)\cdot\left[\frac{\epsilon}{2}+(1-\epsilon)\cdot\Theta(\Delta\bar{Q}_{d:}(\tau))\right]\cdot\left[\frac{(N-1)\cdot\bar{Q}_{dd}(\tau)+(1-\alpha)\cdot\bar{Q}_{dd}(\tau)+\alpha\cdot(\gamma\cdot{Q}^{\max}_{da^{\prime}}(\tau)+r_{d}(\tau))}{N}\right]\\
&+\left\{1-\rho_{d}(\tau)\cdot\left[\frac{\epsilon}{2}+(1-\epsilon)\cdot\Theta(\Delta\bar{Q}_{d:}(\tau))\right]\right\}\cdot\bar{Q}_{dd}(\tau)\\
\bar{Q}_{cd}(\tau+1)=&\rho_c(\tau)\cdot\left[\frac{\epsilon}{2}+(1-\epsilon)\cdot\Theta(-\Delta\bar{Q}_{c:}(\tau))\right]\cdot\left[\frac{(N-1)\cdot\bar{Q}_{cd}(\tau)+(1-\alpha)\cdot\bar{Q}_{cd}(\tau)+\alpha\cdot(\gamma\cdot\bar{Q}^{\max}_{da^{\prime}}(\tau)+r_{d}(\tau))}{N}\right]\\
&+\left\{1-\rho_{c}(\tau)\cdot\left[\frac{\epsilon}{2}+(1-\epsilon)\cdot\Theta(-\Delta\bar{Q}_{c:}(\tau))\right]\right\}\cdot\bar{Q}_{cd}(\tau)\\
\bar{Q}_{dc}(\tau+1)=&\rho_d(\tau)\cdot\left[\frac{\epsilon}{2}+(1-\epsilon)\cdot\Theta(-\Delta\bar{Q}_{d:}(\tau))\right]\cdot\left[\frac{(N-1)\cdot\bar{Q}_{dc}(\tau)+(1-\alpha)\cdot\bar{Q}_{dc}(\tau)+\alpha\cdot(\gamma\cdot\bar{Q}^{\max}_{ca^{\prime}}(\tau)+r_{c}(\tau))}{N}\right]\\
&+\left\{1-\rho_{c}(\tau)\cdot\left[\frac{\epsilon}{2}+(1-\epsilon)\cdot\Theta(-\Delta\bar{Q}_{d:}(\tau))\right]\right\}\cdot\bar{Q}_{dc}(\tau)
\end{array}
\right.
\end{eqnarray}
Here, $\Theta(\cdots)$ is the Heaviside function
\begin{eqnarray}
\Theta(x)=
\left\{
\begin{array}{l}
1, \quad x>0,\\
1/2, \quad x=0,\\
0, \quad x<0
\end{array}\nonumber
\right.
\end{eqnarray}
and $\Delta \bar{Q}_{c:}(\tau)=\bar{Q}_{cc}(\tau)-\bar{Q}_{cd}(\tau)$, $\Delta \bar{Q}_{d:}=\bar{Q}_{dd}(\tau)-\bar{Q}_{dc}(\tau)$.
\rightsep
\begin{multicols}{1}
% Bla bla...
\end{multicols}
\end{figure*}
The final cooperation preference, $\bar{\bf Q}$ and the ranking of convergence rates in the calculation are identical to the simulation results. However, the times series are different during the transient process. The results imply that, on one hand, the mean-field method successfully captures long-term dynamics in the PD RLEGs since all agents' Q-table are identical in the end; on the other hand, the heterogeneity of Q-table for different agents cannot be omitted during transient process and will cause deviations as shown. The numerical results of MS and SD RLEGs also indicate that the mean-field results is decent only if heterogeneity of agents' Q-table is negligible during whole process ( Fig.~S8 and Fig.~S9 in S2.2).

\subsection{Oscillation in the RLEGs for Rock-Paper-Scissors game}

Different from low-degree of freedom of fluctuation for the $2\times 2$ games setting, we pay
attention to the game setting with high-degree of freedom of fluctuation at the equilibrium point, ${\bf p}^{*}$. Here, the RLEGs for rock-paper-scissors (RPS) game is investigated by simulations. In the corresponding RLEGs, the action and state sets are $\mathscr{A}\!=\!\mathscr{S}\!=\!\{R, P, S\}$, and the payoff matrix is
\begin{eqnarray}
\Pi=\left(
\begin{array}{ccc}
\Pi_{rr}& \Pi_{rp} &\Pi_{rs}\\
\Pi_{pr}& \Pi_{pp} &\Pi_{ps}\\
\Pi_{sr}& \Pi_{sp} &\Pi_{ss}
\end{array}
\right)=\left(
\begin{array}{ccc}
0 & -b_{1} & 1 \\
1 & 0 & -b_{1} \\
-b_{1} & 1 & 0
\end{array}
\right)
\end{eqnarray}
with a tunable parameter $b_{1}$. In the traditional EGs, the evolution of collective behaviors is described by ${\bf f}=\{f_{a_{\underline i}}, a_{\underline i}\in\mathscr{A}\}$ in a three-dimensional simplex. Here, $f_{a_{\underline i}}$ is the action preference of $a_{\underline i}$. The mean field treatment~\cite{N:2006} by the RDE of traditional EGs reveals that: 1) the fixed point ${\bf f^*}=\{1/3, 1/3,1/3\}$ is globally stable in the case of $b_{1}< 1$; 2) the fixed point is unstable if $b_{1}>1$; 3) there is a center surrounded by neutral oscillation for $b_{1}=1$. Here, we point out that in the case 2) the trajectories of the replicator dynamics starting from arbitrary interior initial conditions will converge to the boundary of the simplex in oscillations with the increasing amplitude.

\begin{figure}
\centering
\includegraphics[width=0.95\linewidth]{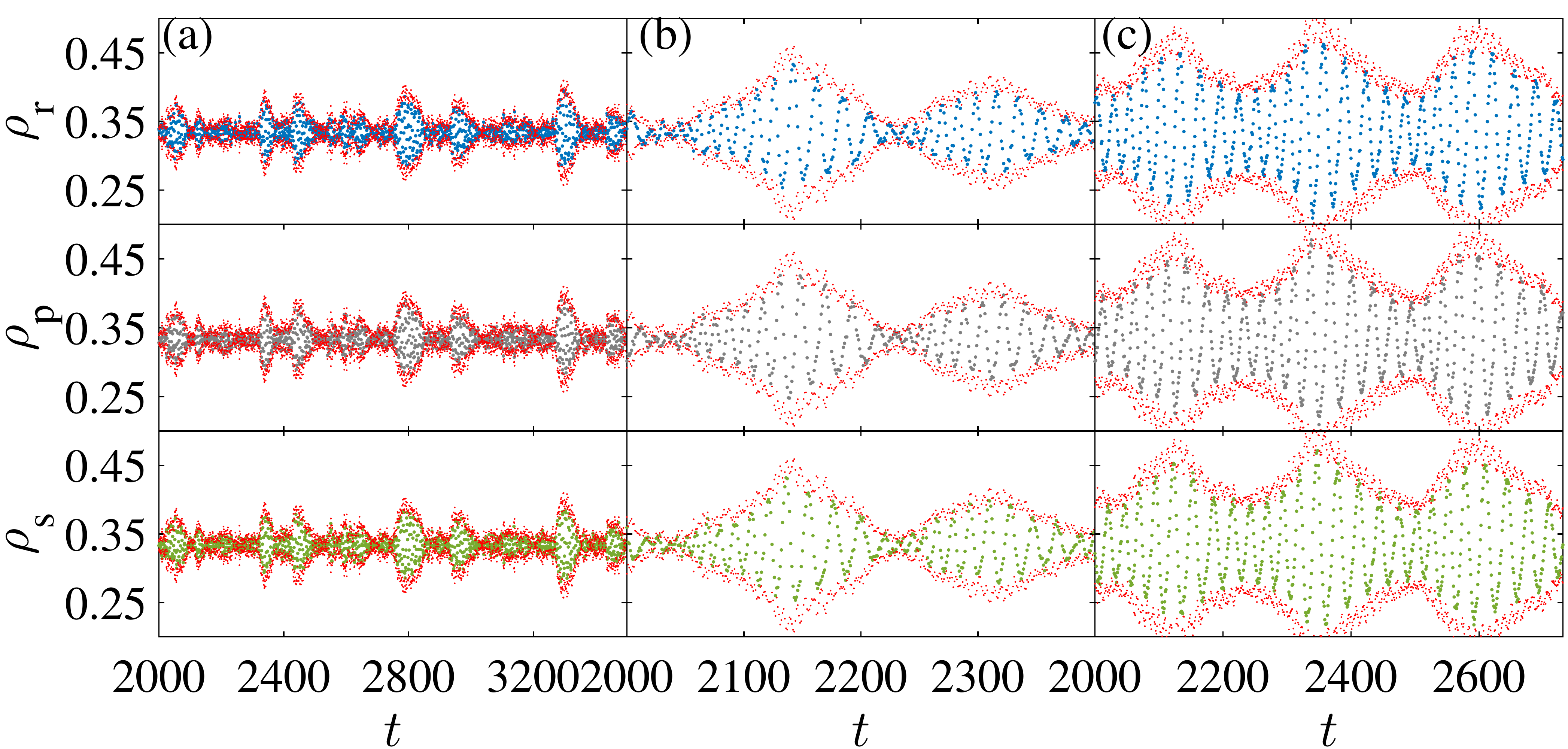}
\caption{ (Color online) {\bf The time series of the three fractions in the RPS RLEGs.}  The parameter in (a-c) $b_{1}=0.7, 1$ and $2$, respectively. The red dots in the figure is $1/3\pm\Delta{\bm\rho}(t)$. The learning parameters are $\alpha=\gamma=0.9$ and $\epsilon=0.02$, with the system size being $N=10000$.}
\label{fig:compare_numerical_simulated}
\end{figure}

Figure~\ref{fig:compare_numerical_simulated} shows the time series of the fraction of the three species in the initiators
\begin{eqnarray}
\rho_{a_{\underline i}}(t)=\frac{\sum_{i=1}^{N}\delta (a(\tau_{t}^{k})-a_{\underline i})}{N}, \forall a_{\underline i}\in{\mathscr A}.
\end{eqnarray}
And the shown with red dots are $1/3\pm\Delta{\bm\rho}$, where $\Delta{\bm\rho}$ is the deviation that is computed as the Euclidean distance between ${\bm\rho}(t)$ and ${\bm \rho}^*$, in which
${\bm \rho}=\{\rho_{a_{\underline i}}|a_{\underline i}\in\mathscr{A}\}$ and ${\bm \rho}^*=\{1/3, 1/3, 1/3\}$. Fig.~\ref{fig:compare_numerical_simulated}(a) shows the case for $b_{1}<1$ where ${\bm\rho}^*$ is globally stable at large except for some small irregular bursts. However,
 in the case of $b_{1}\geq1$ as shown in Fig.~\ref{fig:compare_numerical_simulated}(b, c), clear oscillatory structures emerge in the time series in the RPS RLEGs. The difference for (b) and (c) lies in the envelope of these oscillatory trajectories, where in the case of $b_{1}=1$ there is some moments that the amplitude of oscillation almost diminishes while the oscillation is always significant in the case of $b_{1}>1$.

In the RLEGs for the RPS game, the equilibrium point for shares ${\bf p}^*=\{1/3, 1/3, 1/3\}$, which is mixed and weak. The previous analysis shows that the initiators exploring the optimal mode leads ${\bf p}$ toward to ${\bf p}^*$ and reduce the reward gap between actions gradually. During the exploring process, the amplitude of the oscillation is diminished. However, the agent's belief becomes fragile to fluctuations since the gaps between elements in state row of Q-table are reduced.  ${\bf p}^*$ is stable if the system is able to resist various fluctuations. The simulation shows that the decrease of $b_{1}$ makes the system less sensitive to the fluctuations (a-c) as in EGs. Compare to the EGs, the stability of ${\bf p}^*$ depends more on properties of payoff matrix in RLEGs, they cause abundant collective behaviors in the RLEGs for the identical RPS game setting. Furthermore, the trajectories here are different from the cases in the SD RLEGs because the system needs to resist more comlex and diverse fluctuations to keep ${\bf p}$ stay around ${\bf p}^{*}$.

\section{Discussion and Conclusion}
\label{sec:summary}

In this work, we propose a reinforcement learning evolutionary game model for a general two-player multi-actions ($2\times n_{a}$) game setting, where agents improve their adaption through Q-learning algorithm rather than birth-death process or imitation in the traditional evolutionary games. A striking finding is that the resulting cooperation preference in RLEGs is the same as the traditional EGs in $2\times2$ cases, resting on the Nash Equilibrium of the games. Furthermore we show that Snowdrift game cases present an oscillatory evolution with explosive cooperation in between, exhibiting quite different dynamical mechanism from the traditional EGs.

We investigate dynamics of an analogous individual's Q-table as well as its behavior in a static environment in theory, and apply these to analyze the stability of the RLEGs for the standard $2\times 2$ game setting. The analysis indicates that the  pure and strict equilibrium point is stable in the RLEGs for the game. Different from the stable mixed and weak equilibrium point for the MS region, oscillatory dynamics potentially emerges for the SD region. The qualitative analysis explicitly shows the impact of learning parameters on the oscillatory dynamics in the SD RLEGs: 1) Low learning rate and long-term sight is beneficial to amplitude and period by the increasing cascading effects after the breakdown of stability; 2) The long-sighted agents prolong the delay time to explore optimal action, the quiescent stage, and amplify the amplitude; 3) Higher exploration rate signifies a sharper increase of cooperation preference after the quiescent stage. Therefore, low learning rate and high discount factor as well as exploration rate facilitate the emergence of oscillation in the RLEGs.

Based on these analysis, one learns that a variety of collective behavior results from the dependence of stability at the equilibrium point for shares in RLEGs on more properties of the payoff matrix than in EGs. Furthermore, a numerical method combining with mean-field is put forward to obtain the dynamics of cooperation preference and average Q-table in the PD game. The result is partly consistent with the simulation, especially the final preference and average Q-table, and analysis in the static environment. We also show that the oscillatory dynamics is general in the RLEGs when with a mixed and weak equilibrium point for shares, such as Rock-Paper-Scissors game, where the collective behavior is also quite different from the corresponding traditional EGs.

Our work could lay out a foundation for further systematic investigation of evolutionary games from the perspective of machine learning. First, the simulations and analysis in the static environment contribute to the further understanding analytically in the future. Moreover, the numerical method embedding the mean-field takes the first step to build a mathematical framework of the RLEGs. At last, our work may also aid to understand the explosive events and oscillating behaviors in the society since reinforcement learning actually mimics the introspectiveness in human behavior patterns.

\begin{acknowledgements}
 Li Chen is supported by the National Natural Science Foundation of China under Grants No. 61703257.
 \\
 \\
\textbf{Compliance with ethical standards}
\\
\\
\textbf{Conflicts of interest} The authors declare that they have no conflict
of interest.
\end{acknowledgements}

\bibliographystyle{spphys}
\bibliography{OPCS}
\end{document}

\title{The supplementary}
\author{Si-Ping Zhang   \and
        Ji-Qiang Zhang  \and
        Li Chen   \and
        Xu-Dong Liu
       }
\institute{
           Si-Ping Zhang \at
           The Key Laboratory of Biomedical Information Engineering of Ministry of Education, The Key Laboratory of Neuro-informatics \& Rehabilitation Engineering of Ministry of Civil Affairs, and Institute of Health and Rehabilitation Science, School of Life Science and Technology, Xi'an Jiaotong University, Xi'an 710049, China
           \and
           Ji-Qiang Zhang \at
           Beijing Advanced Innovation Center for Big Data and Brain Computing, Beihang University, Beijing, 100191, China\\
           zhangjq13@lzu.edu.cn
           \and
           Li Chen \at
           School of Physics and Information Technology, Shaanxi Normal University, Xi'an, 710062, China\\
           \and
           Xu-Dong Liu \at
           Beijing Advanced Innovation Center for Big Data and Brain Computing, Beihang University, Beijing, 100191, China\\
           }
\maketitle
% \begin{abstract}
% Given sparse and noisy movement data from a pedestrian crowd - a class of
% active body systems, is it possible to uncover the hidden group interaction
% patterns or connections? Yes, it is possible. Here we develop a general
% framework based on an optimal combination of compressive sensing ($L_1$
% minimization) and the conventional $L_2$ optimization procedure to
% accurately detect the contact network embedded in the pedestrian crowd. The
% optimal weights of the $L_1$ and $L_2$ components can be determined solely
% from data through the basic optimization principle. To detect hidden
% interaction patterns from spatiotemporal data has broader applications, and
% our optimized compressive sensing based framework provides a practically
% viable solution.
% \keywords{Active body system \and Compressive sensing \and $L_1$-regularized least squares \and Optimal detection}
% % \PACS{PACS code1 \and PACS code2 \and more}
% % \subclass{MSC code1 \and MSC code2 \and more}
% \end{abstract}

\section{The reinforce learning in the static environment.} \label{analyis_static}

\subsection{The preliminary of Q-learning for $2$-actions and $2$-states setting}
\begin{figure}[htbp]
\centering
\includegraphics[width=0.8\linewidth]{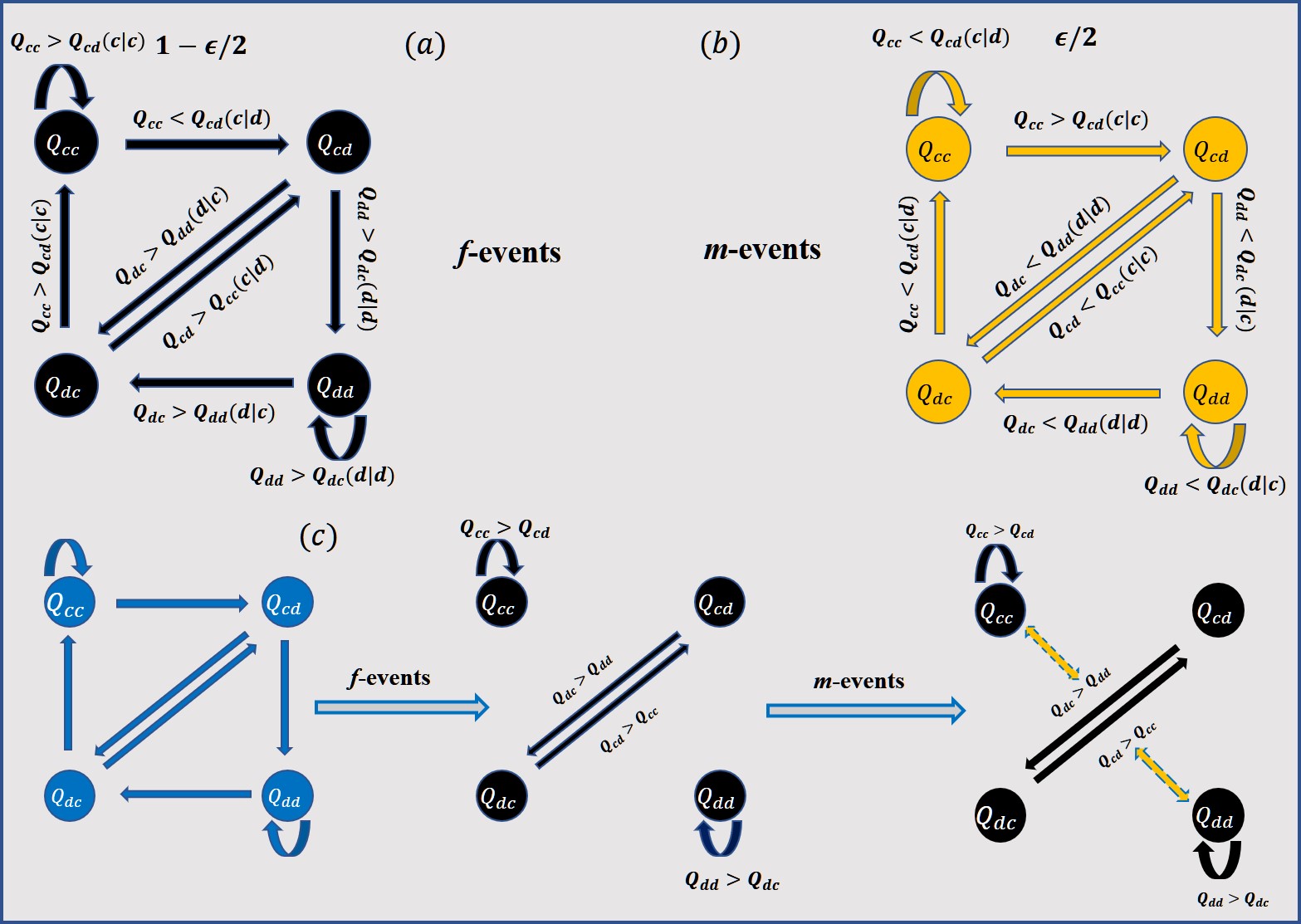}
\caption{ (Color online) {\bf The update hop of elements on Q-table in different events as well as the freezing and melting process for behavior modes.}
(a) and (b) show the hops between elements in Q-table in {\em f}-events and {\em m}-events, respectively. The demands of different hop in (a) and (b) is manifested on the corresponding arrow. (c) shows the freezing and melting process for behavior modes in {\em f}-events and {\em m}-events, respectively. Agents' behavior will be frozen at \uppercase\expandafter{\romannumeral1}) CCM, \uppercase\expandafter{\romannumeral2}) CDM or \uppercase\expandafter{\romannumeral3}) DDM in {\em f}-events. These modes are potentially melted and transformed to each other in {\em m}-events.}
\label{fig:Q-table_update}
\end{figure}

To understand $2\times 2$ RLEGs, we investigate individuals' learning dynamics and behavior features in a static environment. The system consists of a number of non-interacting individuals owning actions set $\mathscr{A}=\{C, D\}$ and state set $\mathscr{S}=\{C, D\}$. However, different from RLEG, $r_{c}$ and $r_{d}$ for actions $C$ and $D$ are constant and the environment is static for each individual. The learning algorithm for individuals is same as for initiators in RLEG. Here, we investigate dynamics of an arbitrary individual's behavior and Q-table since non-interacting individuals are identical in the system. Similar to initiators' update method in RLEG, $i$'s state $s(\tau)$ is replaced previous action $a(\tau-1)$ in the protocol. Then, update hops between the elements in Q-table are described by a double-layers graph as Fig.~\ref{fig:Q-table_update} (a-b) shows. Each hop in Fig.~\ref{fig:Q-table_update} is described via an iteration equation, e.g. the update as  $s(\tau)=C$, $a(\tau)=D$ and $\max \{Q_{da^{\prime}}(\tau)\}=Q_{dd}(\tau)$ is
\begin{eqnarray}\label{eq:fixed_mode1}
&\left(
\begin{array}{c}
Q_{cc}(\tau+1)\\
Q_{cd}(\tau+1)\\
Q_{dc}(\tau+1)\\
Q_{dd}(\tau+1)
\end{array}
\right)=
\left(
\begin{array}{cccc}
1 & 0 & 0 &0\\
0 & 1-\alpha & 0 &\alpha\gamma\\
0 & 0 & 1 &0\\
0 & 0 & 0 &1
\end{array}
\right)\left(
\begin{array}{c}
Q_{cc}(\tau)\\
Q_{cd}(\tau)\\
Q_{dc}(\tau)\\
Q_{dd}(\tau)
\end{array}
\right)+
\alpha\left(
\begin{array}{c}
0\\
 r_{d}\\
0\\
0
\end{array}
\right).\nonumber
\end{eqnarray}
The equation shows the update of elements in Q-table involve the self-coupling (memory effects), inter-elements coupling (effect of future reward), as well as coupling with environment (effect of reward) through learning rate $\alpha$, discount rate $\gamma$ and rewards, respectively. For instance, high $\gamma$ will enhance the effect of future reward but weaken the sensitive to reward for a given $\alpha$.

The learning algorithm in the main text shows individuals have two methods to update action, 1) according to $h$ function or 2) randomly. However, the random action maybe same or opposite to the action as follow $h$-function with an identical probability $\epsilon/2$.
So, we divide update events into two kinds on the basis of individual's action as Fig.~\ref{fig:Q-table_update} (a) and (b) show. Fig.~(a) shows the update events that the individual's action is in line with $h$ function. Fig.~(b) shows opposite cases resulting from exploration events. The update events in Fig.~(a) and (b) are called as {\it freezing events} ({\em f}-events) and {\it melting events} ({\em m}-events), respectively.
The frequency of {\em f}-events $1-\epsilon/2$ is much higher than $\epsilon/2$ in the model.
Therefore, {\em m}-events can be regarded as a disturbances in the world of {\em f}-events.
In addition, we just show the elements associating with events along an update path since no more than two elements in Q-table associate with an update event only.

If only {\em f}-events, each individual's Q-table will be ``{\em frozen}'' in such a static environment finally. When frozen, individuals' behavior will be in one of three modes, i.e. update events form a closed path in the layer of {\em f}-events (Fig.~\ref{fig:Q-table_update}(c) shows)\\
\uppercase\expandafter{\romannumeral1}) frozen cooperation in the form of C-C mode (CCM)
when $\arg \max\limits_{a^{\prime}}\{Q_{sa^{\prime}}\}=C$ for $s=C$;\\
\uppercase\expandafter{\romannumeral2}) frozen defection in the form of D-D mode (DDM) when $\arg \max\limits_{a^{\prime}}\{Q_{sa^{\prime}}\}=D$ for $s=D$;\\
\uppercase\expandafter{\romannumeral3}) cyclic frozen mode in the form of C-D mode (CDM)
when $\arg \max\limits_{a^{\prime}}\{Q_{sa^{\prime}}\}=D$ for $s=C$,
and $\arg \max\limits_{a^{\prime}}\{Q_{sa^{\prime}}\}=C$ for $s=D$. \\
However, when {\em m}-events are present, they not only perturb individual's actions but may also ``{\it melt}'' modes as Fig.~\ref{fig:Q-table_update}(c) shows. Here, we appoint an individual is in a certain mode if the individual is unable to leave it in {\em f}-events, and the individual is at the frozen point of the mode when Q-table is constant in {\em f}-events.
% We call
% these potential executed motifs at any fixed motif set as individuals' behavior ``mode''.

\subsection{The freezing process and behavior modes}

As the individual is in the CCM or DDM at $\tau$ moment, the update of Q-table in {\em f}-events only are that
\begin{eqnarray}
Q_{cc}(\tau^{\prime}+1)=(1-\alpha)Q_{cc}(\tau^{\prime})+\alpha\left[\gamma Q_{cc}(\tau^{\prime}) + r_{c}\right],
\end{eqnarray}
and
\begin{eqnarray}
Q_{dd}(\tau^{\prime}+1)=(1-\alpha)Q_{dd}(\tau^{\prime})+\alpha\left[\gamma Q_{dd}(\tau^{\prime}) + r_{d}\right],
\end{eqnarray}
respectively. Here, $\tau^{\prime}=\tau+n(n\in{\mathbb N})$ denotes time in the CCM or DDM.
The {\em key element} at the frozen point in CCM and DDM are $\max\{Q_{ca^{\prime}}\}=Q_{cc}^{_\text{CCM}}=\frac{r_{c}}{1-\gamma}$ and $\max\{Q_{da^{\prime}}\}=Q_{dd}^{_\text{DDM}}=\frac{r_{d}}{1-\gamma}$, respectively.
The freezing rate are
\begin{eqnarray}
\lambda_{f_{\text{CCM}}}=\frac{|Q_{cc}(\tau^{\prime}+1)-Q_{cc}^{_\text{CCM}}|}{|Q_{cc}(\tau^{\prime})-Q_{cc}^{_{\text{CCM}}}|}=1-\alpha+\alpha\gamma\nonumber
\end{eqnarray}
and
\begin{eqnarray}
\lambda_{f_{\text{DDM}}}=\frac{|Q_{dd}(\tau^{\prime}+1)-Q_{dd}^{_\text{CCM}}|}{|Q_{cc}(\tau^{\prime})-Q_{dd}^{_{\text{DDM}}}|}=1-\alpha+\alpha\gamma,\nonumber
\end{eqnarray}
which suggest high $\alpha$ (short memory) and low $\gamma$ (short-sighted) accelerates the freezing process.

After the individual is in the CDM at $\tau$, updates involve two elements in Q-table under {\em f}-events only, $Q_{cd}$ and $Q_{dc}$. Without loss of generality, we assume $Q_{cd}(\tau)>Q_{cc}(\tau)$ when the individual enter the CDM. Then, updates of Q-table in {\em f}-events only are
% \begin{eqnarray}
% \left(
% \begin{array}{c}
% Q_{cd}(\tau+1)\\
% Q_{dc}(\tau+2)
% \end{array}
% \right)=
% \left(
% \begin{array}{cc}
% 1-\alpha & \alpha\gamma\\
% \alpha\gamma & 1- \alpha\gamma\\
% \end{array}
% \right)
% \left(
% \begin{array}{c}
% Q_{cd}(\tau)\\
% Q_{dc}(\tau+1)
% \end{array}
% \right)
% \end{eqnarray}
\begin{eqnarray}
\left\{
\begin{array}{l}
Q_{cd}(\tau^{\prime}+1)=(1-\alpha)Q_{cd}(\tau^{\prime})+\alpha\left[\gamma Q_{dc}(\tau^{\prime})+r_{d}\right]\nonumber\\
Q_{dc}(\tau^{\prime}+2)=(1-\alpha)Q_{dc}(\tau^{\prime}+1)+\alpha\left[\gamma Q_{cd}(\tau^{\prime}+1)+r_{c}\right]\nonumber
\end{array}
\right.
\end{eqnarray}
% \begin{}
% \begin{eqnarray}\label{eq:cycle_mode}
% \left\{
% \begin{array}{l}
% Q_{cd}(\tau+1)&=&(1-\alpha)Q_{cd}(\tau)\\
% \qquad&+&\alpha\left[\gamma Q_{dc}(\tau)+ r_{d}\right]\nonumber\\
% Q_{dc}(\tau+2)=(1-\alpha)Q_{dc}(\tau+1)\\
% \qquad+\alpha\left[\gamma Q_{cd}(\tau+1)+ r_{c}\right].
% \end{array}
% \right.
% \end{eqnarray}
where $\tau^{\prime}=\tau+2n(n\in \mathbb{N})$ in the CDM.
Since $Q_{dc}(\tau^{\prime})=Q_{dc}(\tau^{\prime}+1)$ and
$Q_{cd}(\tau^{\prime}+1)=Q_{cd}(\tau^{\prime}+2)$, above updates is translated into matrix expression
\begin{eqnarray}
&\left(
\begin{array}{c}
Q_{cd}(\tau^{\prime}+2)\\
Q_{dc}(\tau^{\prime}+2)
\end{array}
\right)&=
\left(
\begin{array}{cc}
1-\alpha& \alpha\gamma\\
\alpha\gamma-\alpha^2\gamma& \quad 1-\alpha+\alpha^2\gamma^2
\end{array}
\right)
\cdot\left(
\begin{array}{c}
Q_{cd}(\tau^{\prime})\\
Q_{dc}(\tau^{\prime})
\end{array}
\right)
+\left(
\begin{array}{c}
\alpha r_{d}\\
\alpha^2\gamma r_{d}+\alpha  r_{c}
\end{array}
\right)
\end{eqnarray}
At the frozen points, the {\em key elements} in CDM are $Q_{cd}^{_\text{CDM}}=\frac{r_{d}+\gamma r_{c}}{1-\gamma^2}$ and
$Q_{dc}^{_\text{CDM}}=\frac{r_{c}+\gamma r_{d}}{1-\gamma^2}$. The freezing rate in two {\em f}-events is
\begin{eqnarray}
\lambda_{f_{\text{CDM}}}=&&\frac{|Q_{cd}(\tau^{\prime}+2)-Q_{cd}^{_\text{CDM}}|}{2(|Q_{cd}(\tau^{\prime})-Q_{cd}^{_{\text{CDM}}}|)}=\frac{|Q_{dc}(\tau^{\prime}+2)-Q_{dc}^{_\text{CDM}}|}{2(|Q_{dc}(\tau^{\prime})-Q_{dc}^{_{\text{CDM}}}|)}\nonumber\\
=&& 2-2\alpha+\alpha^2\gamma^2+\alpha\gamma\sqrt{4-4\alpha+\alpha^2\gamma^2}\nonumber
\end{eqnarray}
which is also positive and negative connected with $\alpha$ and $\gamma$ as in CCM and DDM.
The analysis shows the {\em key elements} at frozen points of three modes are different as
$r_{c}\neq r_{d}$.

\subsection{The melting process and stable mode}

A mode is the {\it stable mode} if the mode is unable to be eroded in {\it m}-events. We insist that there is only one {\it stable} mode as $r_{c}\ne r_{d}$. Then, we investigate the melting process in different modes and assume $r_{c}<r_{d}$ without loss of generality. The {\em key elements} at frozen points in three modes meet $Q_{cc}^{_\text{CCM}}<Q_{dc}^{_\text{CDM}}<Q_{cd}^{_\text{CDM}}<Q_{dd}^{_\text{DDM}}$. The learning algorithm shows elements in a given state row are {\it rivals} because the largest one determine individual's action at the state. In fact, a {\em frozen} mode will lose the stability if one {\em key} element is less than its {\em rival} element in any case. Therefore, we focus the influence of {\em m}-event in various closed path on its {\em rival element}.

\begin{figure}
\centering
\includegraphics[width=0.8\linewidth]{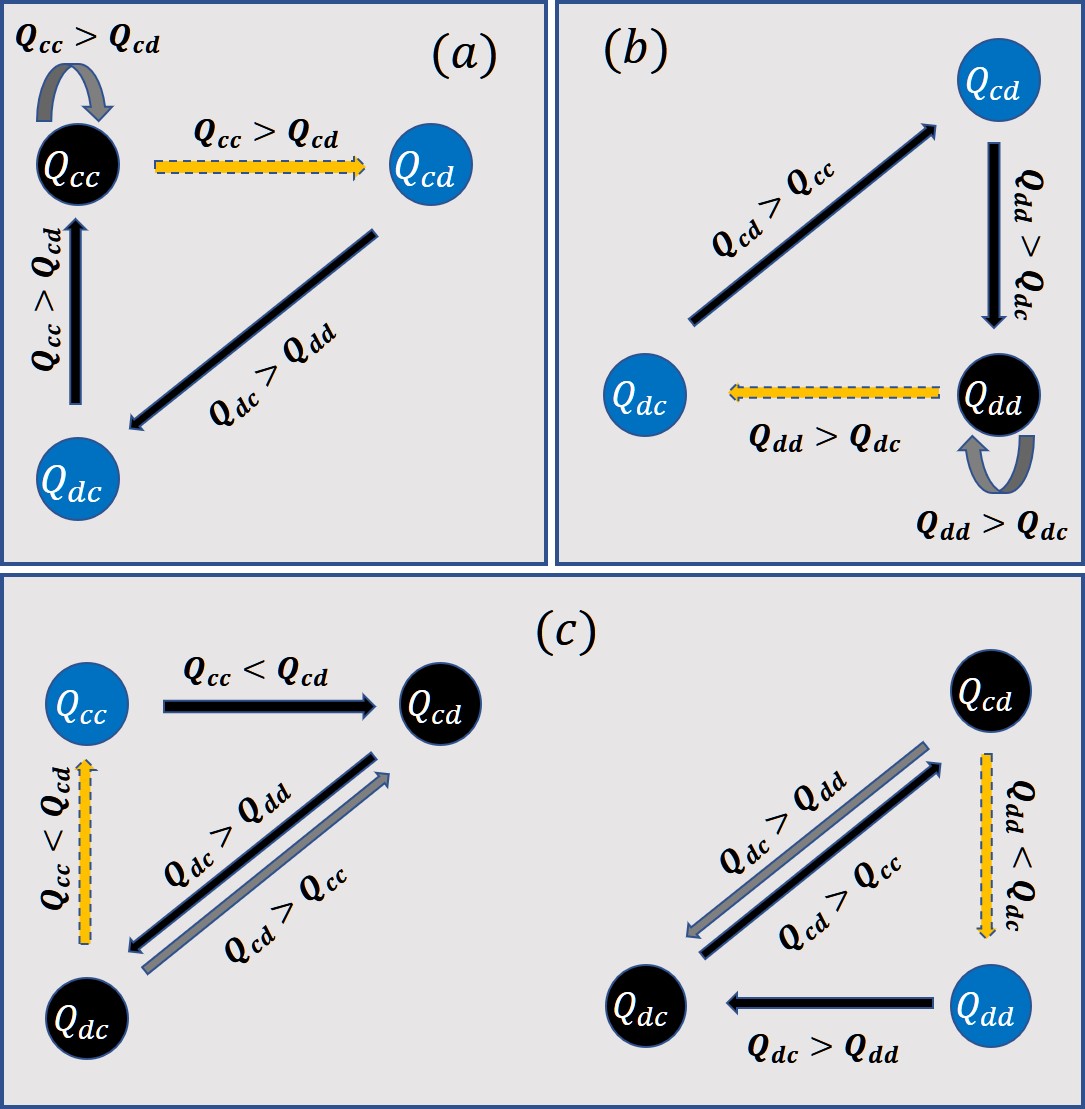}
\caption{ (Color online) {\bf The melting process for different modes in a closed path of events.} (a-b) show the melting process for CCM and DDM in the different path (Eq.(\ref{eq:melting_CCM}) and Eq.(\ref{eq:melting_DDM})). (c) shows the melting process for CDM in two different closed path (Eq.(\ref{eq:melting_CDM1}) and Eq.(\ref{eq:melting_CDM2})). Demands for the hops in the path are showed on the arrows. Here, an gray arrow denotes the hop in {\em f}-event is replaced by one in {\em m}-event (yellow arrows). The black arrows are hops in {\em f}-events.}
\label{fig:melting}
\end{figure}

In the first case, the individual is at the frozen point of CCM, i.e. the {\em key element}, $Q_{cc}=Q_{cc}^{\text{CCM}}$, is constant before the mode is melted. Then, the individual takes action $D$ in an {\it m}-event at $\tau^{\prime}$. The individual is assumed return to CCM in the next two {\em f}-events, i.e. the update events form a closed path (Fig.~\ref{fig:melting}(a)). Then, our attention is payed on the effect of {\em m}-event on the {\em rival} $Q_{cd}$. To distinct {\em f}-events, superscript $\#$ is used to denote update in {\it m}-events. Then, the updates following the path are
\begin{eqnarray}\label{eq:melting_CCM}
&\left(
\begin{array}{c}
Q_{cd}^{\#}(\tau^{\prime}+3)\\
Q_{dc}(\tau^{\prime}+3)\\
Q_{cc}^{_\text{CCM}}(\tau^{\prime}+3)
\end{array}
\right)&=
\left(
\begin{array}{ccc}
1-\alpha & \alpha\gamma & 0\\
0 & 1-\alpha & \alpha\gamma\\
0 & 0 &1-\alpha+\alpha\gamma
\end{array}
\right)\cdot\left(
\begin{array}{c}
Q_{cd}(\tau^{\prime})\\
Q_{dc}(\tau^{\prime})\\
Q_{cc}^{_\text{CCM}}(\tau^{\prime})
\end{array}
\right)+\alpha
\left(
\begin{array}{c}
r_{d}\\
r_{c}\\
r_{c}
\end{array}
\right)
\end{eqnarray}
%%%%%%%%%%%%%%%%%%%
% \begin{subequations}
% \begin{empheq}[left=\empheqlbrace]{align}
% Q_{cd}(\tau^{\prime}+1)&=(1-\alpha)Q_{cd}(\tau^{\prime})\nonumber+\alpha\left[\gamma Q_{dc}(\tau^{\prime})+ r_{d}\right]\nonumber  \\
% Q_{dc}(\tau^{\prime}+2)&=(1-\alpha)Q_{dc}(\tau^{\prime}+1)\nonumber+\alpha\left[\gamma Q_{cc}^{\text{CCM}}(\tau^{\prime}+1)+ r_{c}\right] \nonumber
% \\
% Q_{cc}^{\text{CCM}}(\tau^{\prime}+3)&=(1-\alpha)Q_{cc}^{\text{CCM}}(\tau^{\prime}+2)\nonumber+\alpha\left[\gamma Q_{cc}^{\text{CCM}}(\tau^{\prime}+2)+ r_{c}\right]
% \end{empheq}
% \end{subequations}
%%%%%%%%%%%%%%%%%%%%%%%%%%%%%%%
Above expression indicates $Q_{dc}\rightarrow \gamma Q_{cc}^{_\text{CCM}}+r_{c}=Q_{cc}^{_\text{CCM}}$ and $Q_{cd}\rightarrow \gamma Q_{dc}+r_{d}>Q_{cc}^{_\text{CCM}}$ with the updates along the path. The result manifests $Q_{cd}$ will be greater than $Q_{cc}^{\text{CCM}}$ finally, i.e. CCM is to lose the stability under the erosion of {\em m}-event in the close path.

In the second case, the individual is at frozen point of CDM and {\it rival elements} for {\em key elements} are $Q_{cc}$ and $Q_{dd}$. Therefore, the mode is possible eroded if 1) updates of $Q_{cc}$ cause $Q_{cc}>Q_{cd}^{\text{CDM}}$, or 2) updates of $Q_{dd}$ cause $Q_{dd}>Q_{dc}^{\text{CDM}}$ in {\em m}-events. For the case 1), we assume the individual takes action $C$ in state $C$ at $\tau^{\prime}$ in an {\it m}-event. Here, the individual' behavior is also assumed returns to the CDM along a closed path (Fig.~\ref{fig:melting}(c)). Updates in the path can be expresses as
\begin{eqnarray}\label{eq:melting_CDM1}
&\left(
\begin{array}{c}
Q_{cc}^{\#}(\tau^{\prime}+3)\\
Q_{cd}^{_\text{CDM}}(\tau^{\prime}+3)\\
Q_{dc}^{_\text{CDM}}(\tau^{\prime}+3)
\end{array}
\right)&=
\left(
\begin{array}{ccc}
1-\alpha &\alpha\gamma &0 \\
0 & 1-\alpha &\alpha\gamma\\
0 & \alpha\gamma-\alpha^2\gamma & 1-\alpha+\alpha^2\gamma^2
\end{array}
\right)\cdot
\left(
\begin{array}{c}
Q_{cc}(\tau^{\prime})\\Q_{cd}^{_\text{CDM}}(\tau^{\prime})\\Q_{dc}^{_\text{CDM}}(\tau^{\prime})
\end{array}
\right)+
\left(
\begin{array}{c}
\alpha r_{c} \\\alpha r_{d} \\\alpha^2\gamma r_{d}+\alpha r_{c}
\end{array}
\right)
\end{eqnarray}
% \begin{subequations}
% \begin{empheq}[left=\empheqlbrace]{align}
% Q_{cc}(\tau^{\prime}+1)&=(1-\alpha)Q_{cc}(\tau^{\prime})\nonumber+\alpha\left[\gamma Q_{cd}^{\text{CDM}}(\tau^{\prime})+ r_{c}\right]\nonumber  \\
% Q_{cd}^{\text{CDM}}(\tau^{\prime}+1)&=(1-\alpha)Q_{cd}^{\text{CDM}}(\tau^{\prime})\nonumber+\alpha\left[\gamma Q_{dc}^{\text{CDM}}(\tau^{\prime})+ r_{c}\right]\nonumber
% \end{empheq}
% \end{subequations}
Above updates along the path shows the {\em rival element} $Q_{cc}$ tend to $Q_{cd}^{_\text{CDM}}+ r_{c}<Q_{cd}^{_\text{CDM}}$. It means {\em m}-event in the path is unable to melt the mode. For the case 2), the individual's state and action are assumed at $D$ at $\tau^{\prime}$ in an {\em m}-event. Then, updates of Q-table along the closed path as Fig.~\ref{fig:melting}(d) is that
\begin{eqnarray}\label{eq:melting_CDM2}
&\left(
\begin{array}{c}
Q_{dd}^{\#}(\tau^{\prime}+3)\\
Q_{dc}^{_\text{CDM}}(\tau^{\prime}+3)\\
Q_{cd}^{_\text{CDM}}(\tau^{\prime}+3)
\end{array}
\right)&=
\left(
\begin{array}{ccc}
1-\alpha &\alpha\gamma &0 \\
0 & 1-\alpha &\alpha\gamma\\
0 & \alpha\gamma-\alpha^2\gamma & 1-\alpha+\alpha^2\gamma^2
\end{array}
\right)\cdot
\left(
\begin{array}{c}
Q_{dd}(\tau^{\prime})\\
Q_{dc}^{_\text{CDM}}(\tau^{\prime})\\
Q_{cd}^{_\text{CDM}}(\tau^{\prime})
\end{array}
\right)+
\left(
\begin{array}{c}
\alpha r_{d} \\\alpha r_{c} \\\alpha^2\gamma r_{c}+\alpha r_{d}
\end{array}
\right)
\end{eqnarray}
% \begin{subequations}
% \begin{empheq}[left=\empheqlbrace]{align}
% Q_{dd}(\tau^{\prime}+1)&=(1-\alpha)Q_{dd}(\tau^{\prime})\nonumber+\alpha\left[\gamma Q_{dc}^{\text{CDM}}(\tau^{\prime})+ r_{d}\right]\nonumber  \\
% Q_{dc}^{\text{CDM}}(\tau^{\prime}+1)&=(1-\alpha)Q_{dc}^{\text{CDM}}(\tau^{\prime})\nonumber+\alpha\left[\gamma Q_{cd}^{\text{CDM}}(\tau^{\prime})+ r_{c}\right]\nonumber
% \end{empheq}
% \end{subequations}
Repeat the path, the {\em rival element} $Q_{dd}\rightarrow\gamma Q_{dc}^{_\text{CDM}}+ r_{d}>Q_{dc}$. Thus, $Q_{dd}$ will greater than $Q_{dc}^{\text{CDM}}$ finally. It imply the CDM is unstable in front of erosion of {\em m}-event in the path finally.

Above analysis show that {\em key elements} at frozen point in different modes meet $Q_{cc}^{_\text{CCM}}<Q_{cd}^{_\text{CDM}}<Q_{dc}^{_\text{CDM}}<Q_{dd}^{_\text{DDM}}$. Besides, the CCM and CDM is unstable under erosion of {\em m}-events. Therefore, it is reasonable to conclude that the DDM is {\em stable}. Here, we denote the individual's final Q-table in the stable mode as $\tilde{{\bf Q}}$, whose elements $\tilde{Q}_{dd}$ is identical to $Q_{dd}^{_\text{DDM}}$ logically.

When the individual is at frozen point of DDM, the updates of $Q_{cd}$ and $Q_{dc}$ in a closed path with an {\em m}-event (Fig.~\ref{fig:melting} (b)) are
\begin{eqnarray}\label{eq:melting_DDM}
&\left(
\begin{array}{c}
Q_{dc}^{\#}(\tau^{\prime}+3)\\Q_{cd}(\tau^{\prime}+3)\\\tilde{Q}_{dd}(\tau^{\prime}+3)
\end{array}
\right)&=
\left(
\begin{array}{ccc}
1-\alpha & \alpha\gamma & 0\\
0 & 1-\alpha & \alpha\gamma\\
0 & 0 & 1-\alpha+\alpha\gamma
\end{array}
\right)\cdot\left(
\begin{array}{c}
Q_{dc}(\tau^{\prime})\\Q_{cd}(\tau^{\prime})\\\tilde{Q}_{dd}(\tau^{\prime})
\end{array}
\right)+\alpha
\left(
\begin{array}{c}
 r_{c}\\ r_{d}\\ r_{d}
\end{array}
\right)
\end{eqnarray}
% \begin{subequations}
% \begin{empheq}[left=\empheqlbrace]{align}
% Q_{dc}(\tau^{\prime}+1)=(1-\alpha)Q_{dc}(\tau^{\prime})\nonumber+\alpha\left[\gamma Q_{cd}(\tau^{\prime})+ r_{c}\right]\nonumber \\
% Q_{cd}(\tau^{\prime}+2)=(1-\alpha)Q_{cd}(\tau^{\prime}+1)\nonumber+\alpha\left[\gamma Q_{dd}^*+ r_{d}\right]\nonumber
% \end{empheq}
% \end{subequations}
\begin{figure}[htbp]
\centering
\includegraphics[width=0.35\linewidth]{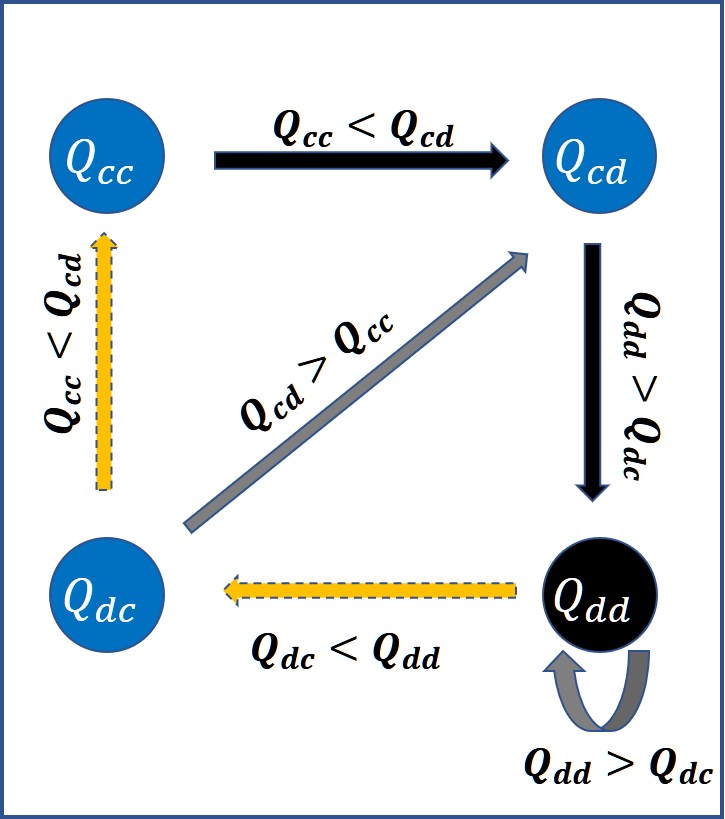}
\caption{ (Color online) {\bf The melting process for DDM in a closed path with consecutive {\em m}-events.} The figure shows melting process for DDM in Eq.~(\ref{eq:melting_CCM}). The expression on arrows is the condition for the hop.
In the figure, an gray arrow denote hops in {\em f}-event is replaced by in {\em m}-events (yellow arrow). Black arrows denote hops in {\em f}-events.}
\label{fig:melting1}
\end{figure}
Then, $Q_{cd}$ and $Q_{dc}$ will converge to $\tilde{Q}_{dd}$ and $\gamma Q_{cd}+ r_{c}$ following $\tilde{Q}_{dd}$ and $Q_{cd}$, respectively. The rate of convergence is
\begin{eqnarray}
\Lambda_{Q_{cd}}=\lim_{\tau\rightarrow \infty}\frac{|Q_{cd}(\tau^{\prime}+3)-\tilde{Q}_{cd}|}{|Q_{cd}(\tau^{\prime})-\tilde{Q}_{cd}|}=1-\alpha \nonumber\\
\Lambda_{Q_{dc}}=\lim_{\tau\rightarrow \infty}\frac{|Q_{dc}(\tau^{\prime}+3)-\tilde{Q}_{dc}|}{|Q_{dc}(\tau^{\prime})-\tilde{Q}_{dc}|}=1-\alpha
\end{eqnarray}
for the {\em m}-event in the corresponding path. The probability of forming the closed path are $\frac{\epsilon}{2}(1-\frac{\epsilon}{2})$. However, the convergence rate in time not only depends on rate of convergence for {\em m}-events but also the probability.

The update of $Q_{cc}$ in DDM demands two consecutive {\it m}-events to form a closed path (Fig.~\ref{fig:melting1}), which has a much lower probability $\frac{\epsilon^2}{2}(1-\epsilon/2)^2$. Updates along the path are
\begin{eqnarray}\label{eq:melting_DDM1}
&\left(
\begin{array}{c}
Q_{dc}^{\#}(\tau+4)\\ Q_{cc}^{\#}(\tau+4)\\
Q_{cd}(\tau+4)\\ \tilde{Q}_{dd}(\tau+4)
\end{array}
\right)&=
\left(
\begin{array}{cccc}
1-\alpha & 0 & \alpha\gamma & 0\\
0 & 1-\alpha &\alpha\gamma &0\\
0 & 0 &1-\alpha &\alpha\gamma\\
0 & 0 & 0 & 1-\alpha+\alpha\gamma
\end{array}
\right)\cdot
\left(
\begin{array}{c}
Q_{dc}(\tau)\\ Q_{cc}(\tau)\\
Q_{cd}(\tau)\\ \tilde{Q}_{dd}(\tau)
\end{array}
\right)+
\alpha\left(
\begin{array}{c}
 r_{c}\\  r_{c}\\
 r_{d}\\  r_{d}
\end{array}
\right)
\end{eqnarray}
% \begin{subequations}
% \begin{empheq}[left=\empheqlbrace]{align}
% Q_{dc}(\tau^{\prime}+1)&=(1-\alpha)Q_{dc}(\tau^{\prime})\nonumber+\alpha\left[\gamma Q_{cd}(\tau^{\prime})+ r_{c}\right]\nonumber \\
% Q_{cc}(\tau^{\prime}+2)&=(1-\alpha)Q_{cc}(\tau^{\prime}+1)\nonumber+\alpha\left[\gamma Q_{cd}(\tau^{\prime})+ r_{d}\right]\nonumber \\
% Q_{cd}(\tau^{\prime}+3)&=(1-\alpha)Q_{cd}(\tau^{\prime}+2)\nonumber+\alpha\left[\gamma Q_{dd}^*+ r_{d}\right]\nonumber
% \end{empheq}
% \end{subequations}
Repeat the path, the element $Q_{cc}$ converge to $\gamma Q_{dc}+ r_{d}$. The rate of convergence for the {\em m}-event in the closed path is
\begin{eqnarray}
\lambda_{m_{\text{cc}}}=\frac{|Q_{cc}(\tau^{\prime}+4)-\tilde{Q}_{cd}|}{|Q_{cd}(\tau^{\prime})-\tilde{Q}_{cd}|}=1-\alpha.
\end{eqnarray}
But, the rate of convergence in time is much lower than $Q_{cd}$ and $Q_{dd}$ since it is difficult to form the path. Here, we point out that $Q_{cd}$ and $Q_{dc}$ also converge to $\tilde{Q}_{dd}$ and $\gamma Q_{cd}+ r_{c}$ along the path.

In sum, $Q_{cc}$, $Q_{cd}$ and $Q_{dc}$ are converge to $\gamma Q_{dc}+ r_{d}$, $\title{Q}_{dd}$ and $\gamma Q_{cd}+ r_{c}$ respectively at last. But, the converge rate of the $Q_{cc}$ much lower than $Q_{cd}$ and $Q_{dc}$ in time as mentioned. According to analysis, we get the final Q-table in the stable mode
\begin{eqnarray}
\tilde{{\bf Q}}=
\left(
\begin{array}{cc}
\frac{\gamma( r_{d}- r_{c})+ r_{c}}{1-\gamma}& \qquad\frac{ r_{d}}{1-\gamma}\\
\frac{\gamma( r_{d}- r_{c})+ r_{c}}{1-\gamma}& \qquad\frac{ r_{d}}{1-\gamma}
\end{array}
\right)
\end{eqnarray}
The final elements in the mode meet $\tilde{Q}_{cd}>Q_{cd}^{_\text{CDM}}$, $\tilde{Q}_{dc}>Q_{dc}^{_\text{CDM}}$ and $\tilde{Q}_{cc}>Q_{cc}^{_\text{CCM}}$. Moreover, the result suggest the mode is stable indirectly because the rival element $Q_{dc}$ is always less than $\tilde{Q}_{dd}$.

\begin{figure}
\centering
\includegraphics[width=1.0\linewidth]{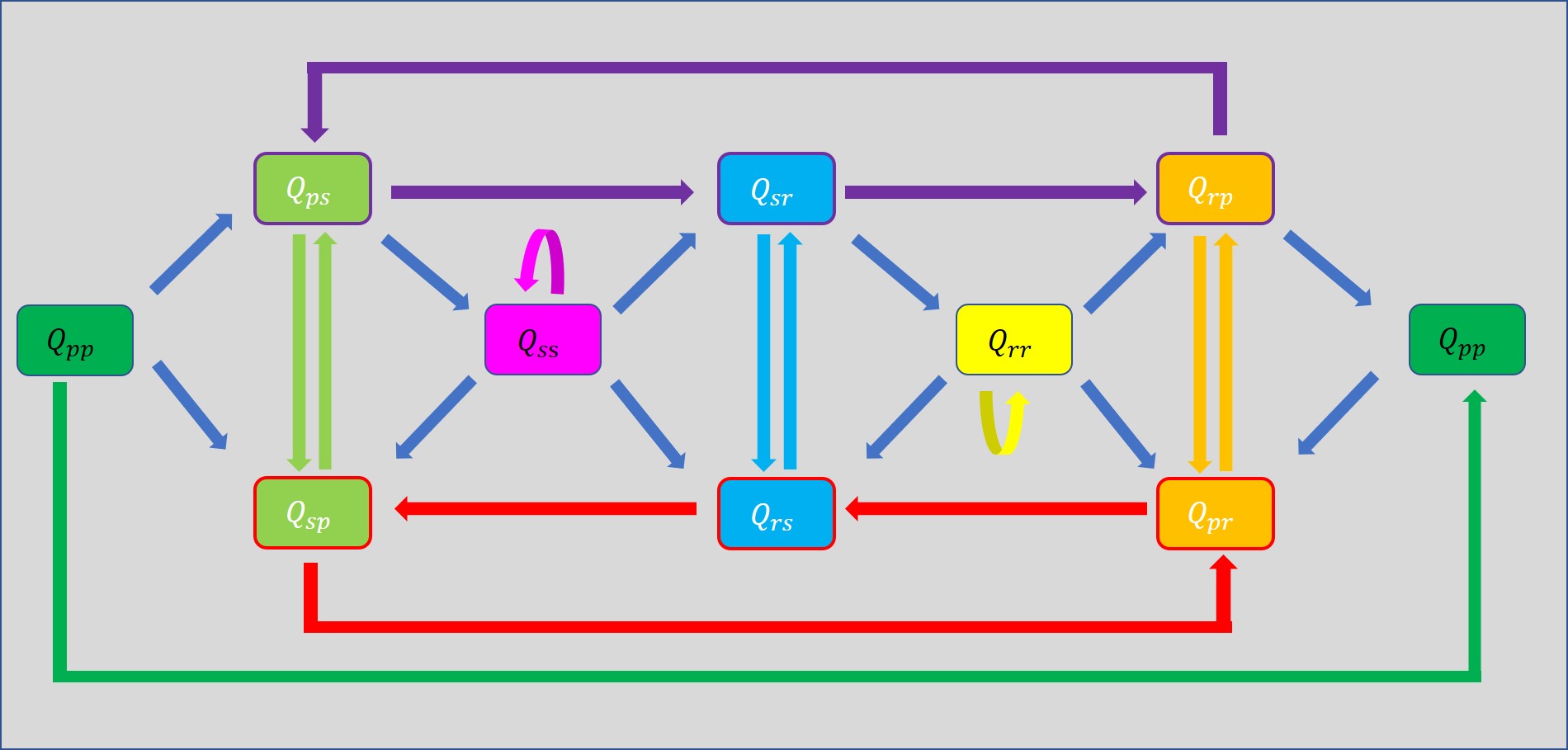}
\caption{ (Color online) {\bf The graph of update hops of Q-table for the RLGs with $3$-actions and $3$-states in {\em f}-events.} Hops and elements in different modes are marked with different colors except the dark blue. Elements associated with two modes are marked with a box with a different color frame, e.g. $Q_{sp}$ is associated with SPRM and SPM. There are eight feasible modes for individuals.}
\label{fig:RPS_hop}
\end{figure}
Above analysis manifest that there are three feasible modes for the individual at the frozen point in the system. However, only one mode is stable confronting the erosion of {\it m}-events as $ r_{c}\ne r_{d}$. Individuals urging by the reward maximization will adjust their Q-table and mode. In the adjustment of Q-table, high $\alpha$ increase the converge rate for each corresponding cyclic sequence of events. In time, the converge rate also rely on the probability of the events.
Moreover, the stability $\text{CCM}<\text{CDM}<\text{DDM}$ if $ r_{c}< r_{d}$, which cause individual's behavior form a mode flow $\text{CCM}\rightarrow\text{CDM}\rightarrow\text{DDM}$ as $\epsilon\rightarrow 0$. At last, we discover that gap between the {\em rival elements} in the stable $\tilde{{\bf Q}}$ will be reduced with the gap $|r_{c}-r_{d}|$, i.e. individuals' mode become fragile with the decrease of rewards gap.

\subsection{The updating graph for $3$-actions and $3$-states}

The modes for $3$-actions and $3$-states setting is much more complex than for $2$-actions and $2$-states setting. For a $3$-actions and $3$-states setting, the actions set and states set are $\mathscr{A}=\mathscr{S}=\{R, P, S\}$. There are eight feasible modes for individuals in the updating graph shown in Fig.~\ref{fig:RPS_hop}. Here, we figure out the modes containing identical actions is probably different, e.g. the mode in the form of cyclic R-S-P (RSPM) is different from in the form of cyclic R-P-S (RPSM). On the basis of the graph, one learns there are more melting channels to erode each {\em frozen} mode compare with $2$-actions and $2$-states setting. Because of the complexity of graph and melting channels, we abandon the subtle analysis on the multi actions setting in a static environment as in the previous and merely provide the updating graph.

\section{Supplementary results in $2\times 2$ RLEGs}
In the section, we provide supplements in $2\times 2$ for the main text and investigate SH RLEGs further in simulation. The supplementary for the M2.2 are manifested in \ref{subsec:SD_RLEGs} and \ref{subsec:SH_RLEGs} mainly. Furthermore, we investigate the SH RLEGs further by simulation in \ref{subsec:SH_RLEGs}. The supplements about SD and MS numerical results for ~M3.3 is provided in \ref{numerical}.

\subsection{The further simulation and investigation in SD RLEGs}\label{subsec:SD_RLEGs}
\begin{figure}
\centering
\includegraphics[width=0.75\linewidth]{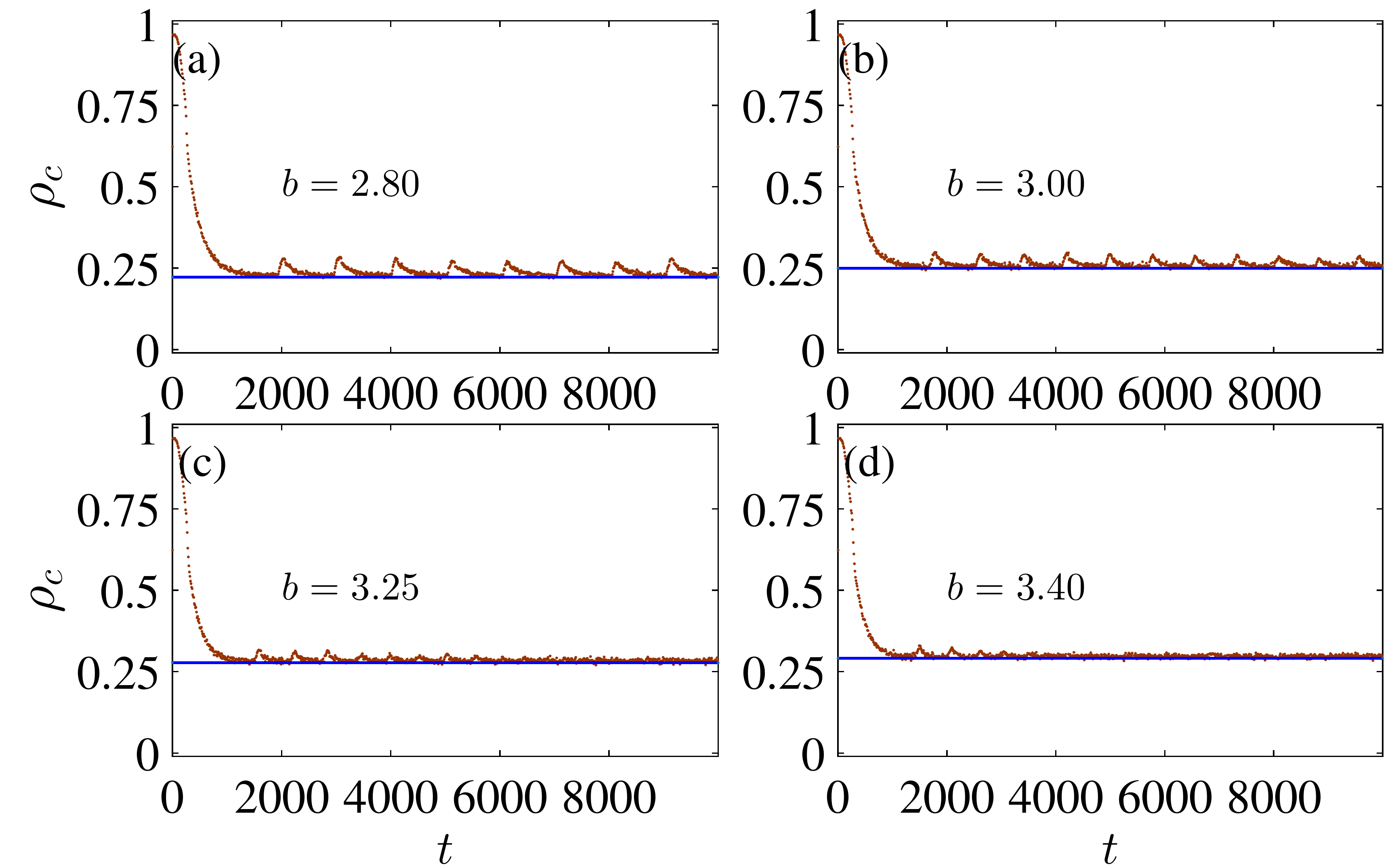}
\caption{ (Color online) {\bf The time series of $\rho_{c}$ over MC step in standard SD RLEGs.} The learning parameters in panels $\alpha=\gamma=0.9$, $\epsilon=0.02$. The number of agents $N=10000$. For learning parameters combination, the transition point between oscillation and non-oscillation is $b^{\prime}=3.24$ as Fig.~M2 shows.}
\label{fig:rhoc-t}
\end{figure}

Fig.~M2 shows there is transition point $b^{\prime}$ for $\langle\rho\rangle$ as the function
of $b$ in the standard SD RLEGs. The cooperation preference is changed abruptly around the point. Thereupon, we conjecture $b^{\prime}$ is the transition point of the collective behavior in the SD RLEGs. As $b<b^{\prime}$, a periodic oscillation emerge in the system. For $b>b^{\prime}$ the oscillation is replaced by a non-oscillation form. Here, we provide a batch of simulation to verify the conjecture in the Fig.~\ref{fig:rhoc-t}. The result manifest the amplitude $A$ is decrease as $b$ close to $b^{\prime}$ gradually. And, the oscillation fades away as $b>b^{\prime}$. The simulation interprets why the cooperation preference is changed fiercely near $b^{\prime}$.

\begin{figure}
\centering
\includegraphics[width=0.75\linewidth]{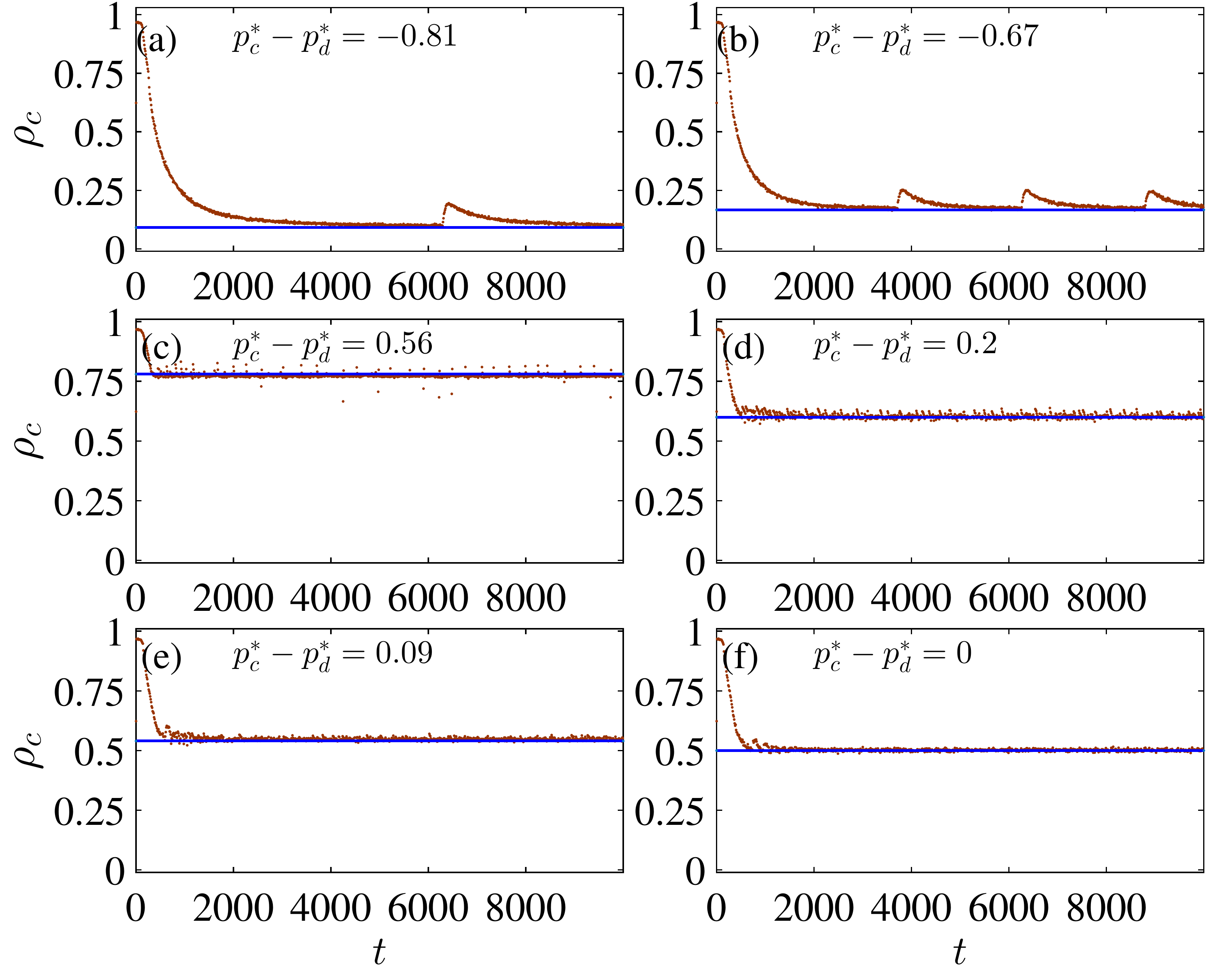}
\caption{ (Color online) {\bf The time series of $\rho_{c}$ over MC step in various SD RLEGs.} The game parameter in (a-f) $b_{2}=2.1, 2.2, 5.5, 3.5, 3.2$ and $3.0$ for $\Pi=(6, b_{2}; 7, 2)$, respectively. The learning parameters $\alpha=\gamma=0.9$, $\epsilon=0.02$ in all panels. The scale of system $N=10000$.}
\label{fig:rhoc-t1}
\end{figure}

Furthermore, the analysis in M3.2 suggest $\Delta \Pi_{:d}/(\Delta \Pi_{:d}+\Delta \Pi_{:c})=1/2$ is the threshold of oscillation form in the SD RLEGs. However, $\Delta \Pi_{:d}/(\Delta \Pi_{:d}+\Delta \Pi_{:c})$ always smaller than $1/2$ in the standard SD RLEGs. Thereby, we replaced the standard form with $\Pi=(6, b_{2}; 7, 2)$, in which $b_{2}$ is also a turning parameter as $b$ in the standard form. The simulation shows a arrhythmic oscillation arises as $p_{c}-p_{d}>0$ (Fig.~\ref{fig:rhoc-t1} (c-d)), which is quite different from the periodic oscillation in (a-b). However, the oscillation fades away as
$p_{c}^*-p_{d}^*$ tend to zero (e-f). Therefore, our analysis in the main text is verified by the simulation.

\subsection{The further investigation on initialization in SH RLEGs}\label{subsec:SH_RLEGs}
In the replicate dynamic equation (RDE) of traditional evolutionary game (EG) for Stag Hunt (SH) game setting, $f_{c}^*=1$ and $f_{c}^*=0$ are bistable fixed points on the cooperation preference. There is an unstable interior fixed point, $f_{c}=\Delta\Pi_{:d}/(\Delta \Pi_{:d}+\Pi_{:c})$. The system will converge to all-cooperation if initial cooperation preference $f_{c_{0}}<\Delta\Pi_{:d}/(\Delta \Pi_{:d}+\Pi_{:c})$, otherwise, defection is dominating. In other words, the convergence direction determined by the initial cooperation preference. Interesting questions are then raised in SH RLEGs: \emph{whether there are bistable fixed cooperation preference? And, whether there is an appropriate Q-table setting under which the convergence direction determined by the initial cooperation preference?}

As a matter of fact, the analysis in M3.2 indicates there are two pure strict equilibrium points and a mixed weak equilibrium point in the SH RLEGs. And, the strict points are stable while the weak point is unstable. Thus, there are bistable fixed cooperation preference as in the EGs in analysis. It is reasonable to assume there is attraction domain for each stable fixed cooperation preference. And the convergence direction depends on initialization locate at which attraction domain in the phase space.

\begin{figure}
\centering
\includegraphics[width=0.75\linewidth]{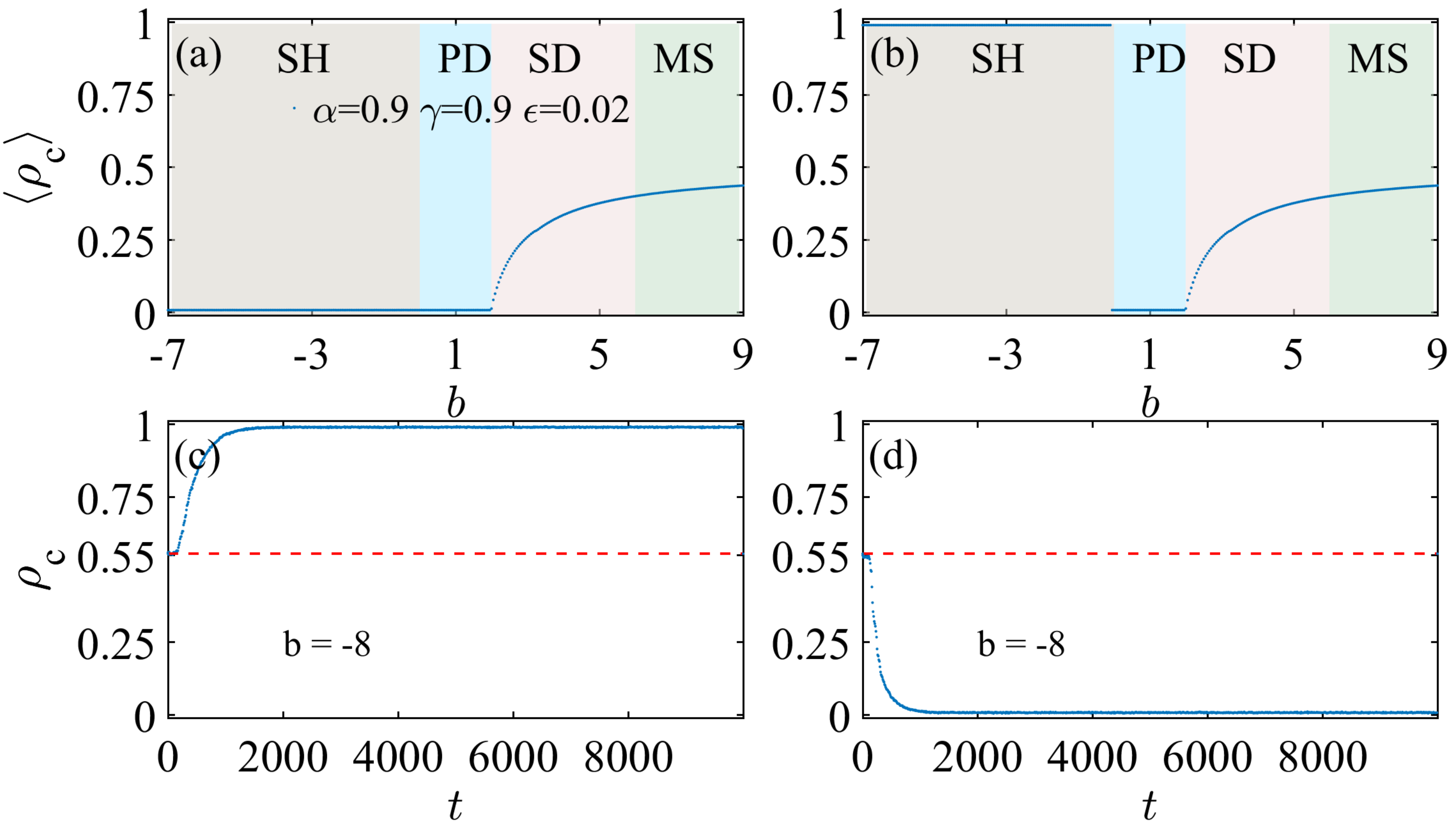}
\caption{ (Color online) {\bf The function of expectation of cooperation preference $\langle\rho_{c}\rangle$ of $b$ and times series of $\rho_{c}$ over MC step in SH RLEGs.} (a-b) shows the function $\langle\rho\rangle$ of $b$ under two initializations. In (a), the initial setting is standard as the main text shows. In (b), initial state is cooperation and Q-table meet $Q_{cc}>Q_{cd}$ and $Q_{dc}>Q_{dd}$ for all agents. (c-d) shows the time series of $\rho_{c}$ in the SH RLEGs under an identical initialization to Q-table but different to $p_{c_{0}}$ (see 2.2). There is unstable interior cooperation preference (dash line) to divide the convergence direction under the initial Q-table setting. In (c-d), the gaming parameter $b=8$, i.e. $p_{c}^*=0.555$, and $p_{c_{0}}=0.56$, $0.55$. The learning parameters in panels $\alpha=\gamma=0.9$, $\epsilon=0.02$. The number of agents $N=10000$.}
\label{fig:SM_rho-b}
\end{figure}

Here, we appoint the initialization in the protocol of main text as the standard initialization.  To survey the influence of initialization on the convergence of cooperation preference, we replace the standard initialized method with a special method in our simulation. Different from the {\em standard initialization} in (a) of Fig.~\ref{fig:SM_rho-b}, the each agent is at state $C$ and its Q-table meets conditions, $Q_{cc}>Q_{cd}$ and $Q_{dc}>Q_{dd}$, in the initialization. The result in (a) and (b) shows there are bistable cooperation preferences in RLEGs for the SH game setting as the analysis in M3.2.

For the high-dimensional RLEGs, the exact attraction domain of each fixed point is hard to be achieved. Less-than-ideal alternative, we focus on how to control the convergence direction in a certain initialize method. In other word, the method is able to insure the initial point in the desired attraction domain. On the basis of results in the static, we explore a method to initialize agents' Q-table, under which the convergence direction is only determined by the $p_{c_{0}}$ as $\epsilon\rightarrow 0$. Here, $p_{c_{0}}$ is the initial fraction of cooperators in the agents following $h$ function.

The analysis in the static shows that a {\em frozen} individual's mode can be only eroded in {\em m}-events even though it is not the optimal. So, we set all agents are in a mode at $p_{c_0}$ and their mode cannot be eroded in {\em f}-events, i.e. almost all agents are frozen in their initial mode before $p_{c_{0}}$ changes for the initialization. In the initialization, 1) $Q_{cc}>Q_{cd}$, $Q_{dc}>Q_{dd}$ for agents in CCM, 2) $Q_{dd}>Q_{dc}$ and $Q_{cd}>Q_{cc}$ for agents in DDM, and 3) $Q_{cd}>Q_{cc}$ and $Q_{dc}>Q_{dd}$ for agents in CDM. Besides, $\max\limits_{sa} \{Q_{sa}\}<\min \{Q_{cc}^{_\text{CCM}}, Q_{cd}^{_\text{CDM}}, Q_{dc}^{_\text{CDM}}, Q_{dd}^{_\text{DDM}}\}$ for all agent. Then, the convergence direction only depends on $p_{c_0}$ that cooperation dominates if $p_{c_{0}}>p_{c}^*$, otherwise defection dominates. The method is employed to initialize agents' Q-table and state in Fig.~\ref{fig:SM_rho-b} (c-d). In (c) and (d), $p_{c_{0}}$ is just slightly greater and less than $p_{c}^*$, respectively. The simulation shows the convergence direction can be controlled via initial $p_{c_{0}}$ in our method.

\subsection{The numerical result in the RLEGs for SD and MS game settings}\label{numerical}

\begin{figure}[htbp]
\centering
\includegraphics[width=0.6\linewidth]{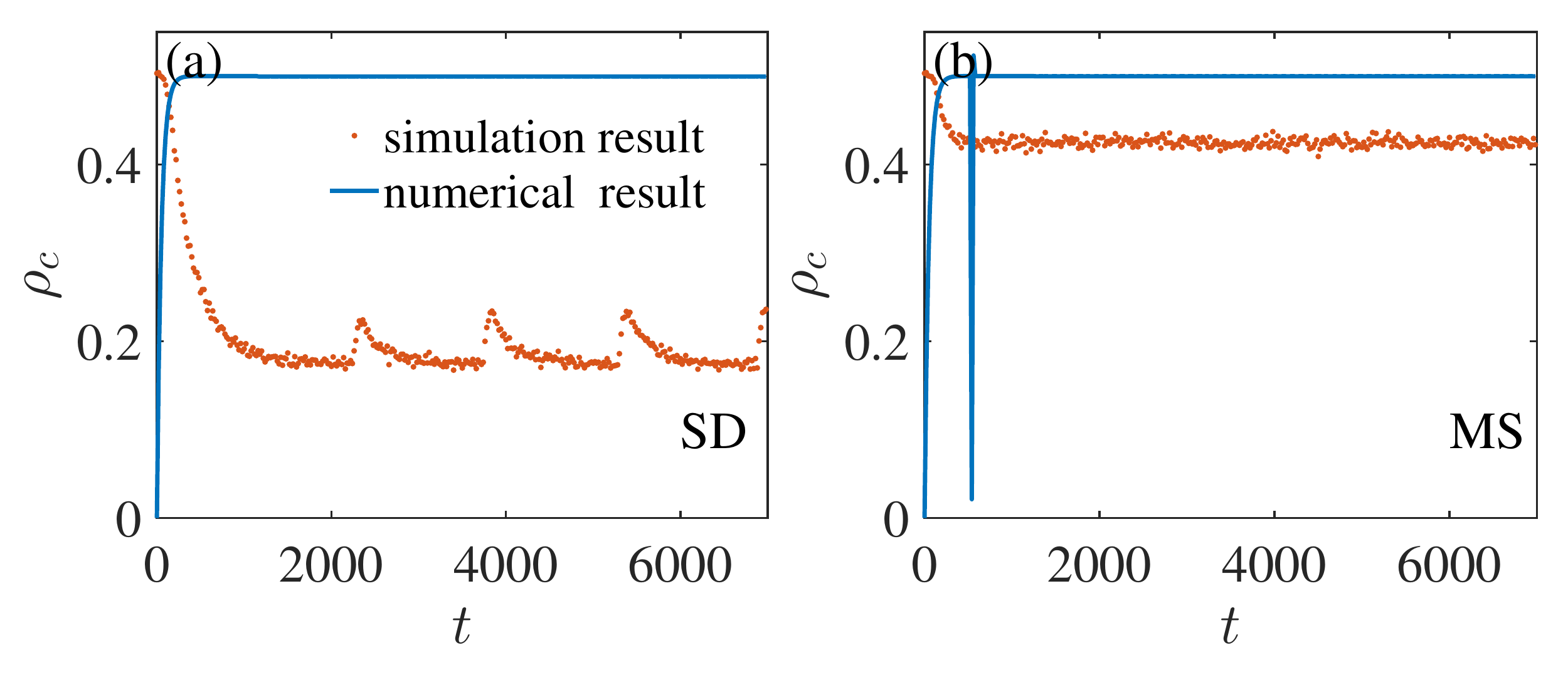}
\caption{ (Color online) {\bf The compare between simulation and numerical result for the times series of $\rho_{c}$ over MC step in the RLEGs with a mixed $p_{c}^*$.} In (a) and (b), the tuning parameters of the standard matrix are $b=7.5$ ($p_{c}^*=0.423$) and $2.5$ ($p_{c}^*=0.1667$) for MS and SD games.
The learning parameters are $\alpha=\gamma=0.9$, $\epsilon=0.02$. The number of agents is $N=10000$.}
\label{fig:sd_ms_rhoc}
\end{figure}

\begin{figure}[htbp]
\centering
\subfigure[SD RLEG]{
\begin{minipage}[t]{0.5\linewidth}
\centering
\includegraphics[width=\linewidth]{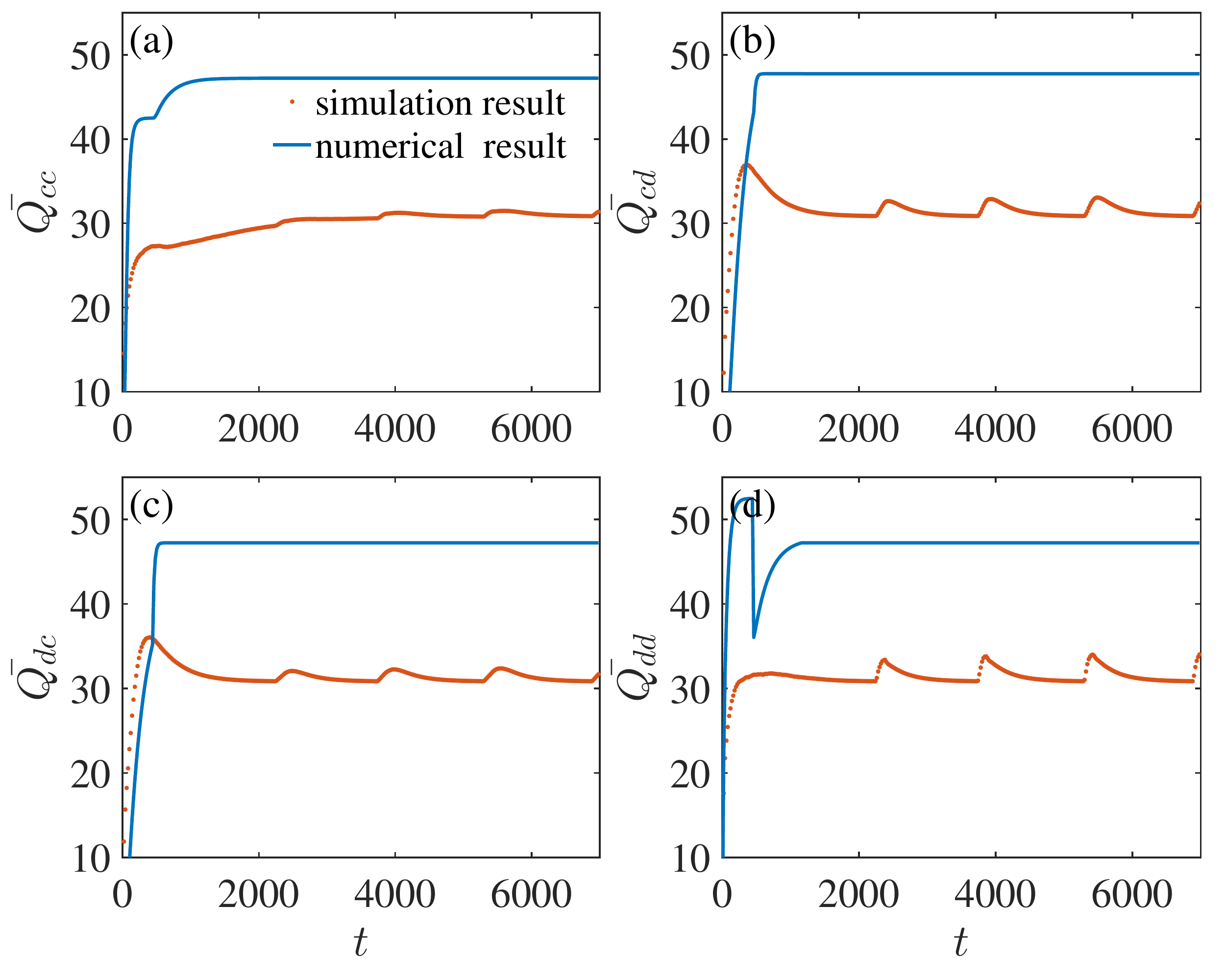}
%\caption{fig1}
\end{minipage}%
}%
\subfigure[MS RLEG]{
\begin{minipage}[t]{0.5\linewidth}
\centering
\includegraphics[width=\linewidth]{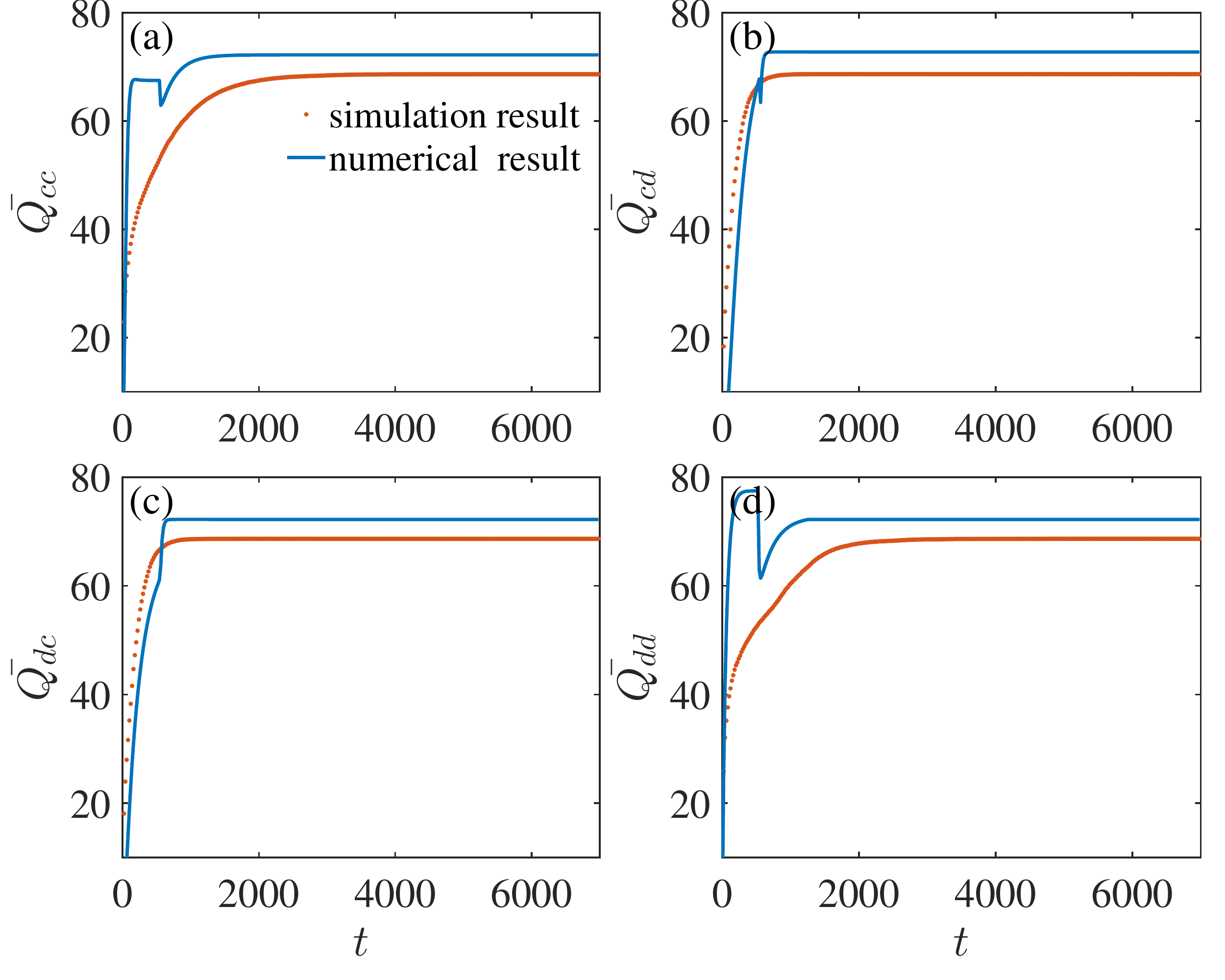}
\end{minipage}%
}
\centering
\caption{ {\bf The compare between simulation and numerical result for the times series of $\bar{Q}_{sa}$ over MC steps in MS and SD RLEGs.} In RLEG for the MS and SD game, the parameters are shared with Fig.~\ref{fig:sd_ms_rhoc} (a) and (b), respectively.}
\label{fig:sd_ms_Q-able}
\end{figure}

In the M3.3, a numerical method embedding mean-field is raised to apprehend RLEGs. In the method, we omit the heterogeneity of agents' Q-table so as to simplify the analysis and employ a homogeneous average Q-table for all agents. The method is decent as agents' Q-table is homogeneous at last, such as PD RLEG with a pure equilibrium point ${\bf p}^*$. It is able to give the final cooperation preference and agents' Q-table. Here, we state that the method is deficient via simulation if the heterogeneity of Q-table for different agents is non-negligible. In the case, the method always prove an identical final cooperation preference $\rho_{c}(\infty)=1/2$ and average Q-table if ${\bf p}^*$ is weak and unique (Fig.~\ref{fig:sd_ms_rhoc} and Fig.~\ref{fig:sd_ms_Q-able}).

\bibliographystyle{spphys}
\bibliography{OPCS}